\documentclass[aps,prc,twocolumn,groupedaddress,nofootinbib]{revtex4-2}
\RequirePackage{plautopatch}
\usepackage{bm}
\usepackage{graphicx}
\usepackage{ascmac}
\usepackage{amsmath,amssymb}
\usepackage{cancel}
\usepackage{url}
\usepackage[setpagesize=false,hidelinks,pdfusetitle]{hyperref}
\usepackage{mathtools}
\usepackage{physics}
\usepackage{physics2}
\usephysicsmodule{braket}
\usephysicsmodule{ab}
\usepackage{derivative}
\usepackage{orcidlink}
\allowdisplaybreaks[1]

\usepackage[normalem]{ulem}  
\usepackage{color} 

\renewcommand\sout{\bgroup \color{red} \ULdepth=-.5ex \ULset}

\begin{document}


\title{Compositness and wave function of shallow bound states\\
in relation to scattering observables}


\author{Ibuki Terashima\orcidlink{0009-0000-9695-4578}}
\email[]{terashima-ibuki@ed.tmu.ac.jp}
\affiliation{Department of Physics, Tokyo Metropolitan University, Hachioji 192-0397, Japan}
\author{Tetsuo Hyodo\orcidlink{0000-0002-4145-9817}}
\email[]{hyodo@rcnp.osaka-u.ac.jp}
\thanks{Present address: Research Center for Nuclear Physics (RCNP), Ibaraki, Osaka 567-0047, Japan}
\affiliation{Department of Physics, Tokyo Metropolitan University, Hachioji 192-0397, Japan}


\date{\today}

\begin{abstract}
We study the internal structure of exotic hadrons, especially focusing on the relation between the compositeness and physical observables. Defined as the probability of finding hadronic molecular components in the wave function, compositeness serves as a quantitative measure of the internal structure of exotic hadrons. We utilize the coupled-channel potential model incorporating both quark and hadron degrees of freedom, which naturally generate the ``bare state'' responsible for the elementary component as the bound state in the quark channel. The behavior of the compositeness under the variation of the model parameters is investigated by using the $X(3872)$ as an example. In particular, we analyze the associated scattering phase shifts and the bound-state wave functions to discuss the relation between the compositeness and the scattering observables for a shallow bound state. As a phenomenological application of the present framework, the compositeness of the $X(3872)$, $T_{cc}(3875)$, $D_{s0}(2317)$, and $D_{s1}(2460)$ is discussed.
\end{abstract}


\maketitle


\section{Introduction}\label{sc_intro}

The recent discoveries of exotic hadrons in the heavy-quark sector, such as the $X(3872)$~\cite{Belle:2003nnu}, the pentaquark states $P_c$~\cite{LHCb:2015yax,LHCb:2019kea}, and the tetraquark candidate $T_{cc}$~\cite{LHCb:2021vvq,LHCb:2021auc}, suggest that the structure of hadrons extends beyond the conventional mesons and baryons~\cite{Hosaka:2016pey,Guo:2017jvc,Brambilla:2019esw,Hyodo:2020czb}. Indeed, a variety of internal structures have been proposed, including multiquark states, hadronic molecules, and gluonic hybrids. Studies of such exotic hadrons are expected to provide crucial insights into how hadrons are formed through the strong interaction.

Among recent approaches to unveiling the internal structure of exotic hadrons, the concept of compositeness has attracted considerable attention~\cite{Weinberg:1965zz,Baru:2003qq,Hyodo:2011qc,Hyodo:2013nka,Kamiya:2015aea,Kamiya:2016oao,Oller:2017alp,vanKolck:2022lqz,Kinugawa:2024crb}. The compositeness is defined as the probability of finding a hadronic molecular component in the wave function of an exotic hadron, and it serves as a useful and quantitative indicator of the internal structure.

The internal structure of exotic hadrons is expected to be ultimately tested through experiments. To achieve this, it is essential to relate the compositeness to experimentally accessible observables. Although the compositeness is not an observable in a strict sense and hence generally exhibits model dependence~\cite{Kinugawa:2024crb}, it is known that, for $s$-wave bound states near the threshold, the compositeness can be related to the scattering length and effective range in a model-independent manner via the weak-binding relations~\cite{Weinberg:1965zz,Kamiya:2015aea,Kamiya:2016oao,Li:2021cue,Albaladejo:2022sux,Kinugawa:2022fzn}. In addition, various relations between the compositeness of weakly bound states and quantities such as the coupling constant and binding energy have been investigated~\cite{Sazdjian:2022kaf,Lebed:2022vks,Kinugawa:2023fbf,Song:2023pdq}.

In Ref.~\cite{Terashima:2023tun}, we have constructed a potential model in which the quark degrees of freedom are coupled with the hadronic degrees of freedom, and have analyzed the structure of the $X(3872)$ as a case study. This model has the advantage of naturally incorporating quark-originated components while also enabling the investigation of the spatial dependence of the wave function through the use of the potential.

In this study, we apply the model proposed in Ref.~\cite{Terashima:2023tun} to investigate the compositeness of the $X(3872)$, which is an $s$-wave weakly bound state. We analyze its dependence on the binding energy, the bare-state energy, the momentum cutoff, and the strength of the interaction in the hadronic channel. Furthermore, we examine the corresponding scattering phase shifts and wave-function behaviors to discuss how differences in compositeness affect various physical quantities. We also assess the validity of the local approximation to the effective potential, which becomes nonlocal upon integrating out the quark degrees of freedom. 

The structure of this paper is as follows. In Sec.~\ref{sec:formulation}, we present the formulation of the coupled-channel model, deriving the bound-state wave function, expression for the compositeness of the bound state, and the scattering amplitude. Section~\ref{sec:results} is devoted to numerical results. We examine how the compositeness, bound-state wave function, and scattering phase shift depend on the model parameters. Finally, we discuss the validity of the local approximation for the effective potential and its relation to the compositeness. The summary of this work is presented in Sec.~\ref{sec:summary}. Preliminary results on the compositeness can be found in a conribution to conference proceedings~\cite{Terashima:2025ksm}.

\section{formulation}\label{sec:formulation}

\subsection{Hamiltonian and wave function}

In this section, we consider the coupled-channel problem of quark and hadronic degrees of freedom within the framework of nonrelativistic quantum mechanics. In addition to the formulation presented in Ref.~\cite{Terashima:2023tun}, we explicitly introduce direct interactions between hadrons. By applying the Feshbach method~\cite{Feshbach:1958nx,Feshbach:1962ut}, we analytically derive the effective interaction between hadrons, in which the quark degrees of freedom are implicitly embedded.

The Hamiltonian $H$, wave function $\ket{\psi}$, and Schr\"odinger equation are given by
\begin{align}
    H
    &=
    \begin{pmatrix}
        T^{q} & 0     \\
        0     & T^{h}
    \end{pmatrix}
    +
    \begin{pmatrix}
        V^{q} & V^t   \\
        V^t   & V^{h}
    \end{pmatrix}, 
    \label{eq:hamiltonian}
    \\ 
    \ket{\Psi}
    &=
    \begin{pmatrix}
        \ket{q} \\
        \ket{h}
    \end{pmatrix}, \\
    H \ket{\Psi} 
    &=  E \ket{\Psi}, 
\end{align}
where $\ket{q}$ is the wave function in the quark channel and $\ket{h}$ is that in the hadron channel, which are coupled to each other through the transition potential $V^t$. $V^q$ is the confining potential in the quark channel, while $V^h$ is the scattering potential in the hadron channel. $T^q$ and $T^h$ are the kinetic energy operators in the quark and hadron channels, respectively. We set the threshold energy of the hadron channel at $E=0$. 

Following Ref.~\cite{Terashima:2023tun}, we derive the coordinate representation of the effective hadron-hadron potential by the Feshbach method as
\begin{align}
    V(\bm{r}',\bm{r},E)
    =\braket[3]{\bm{r}'}{V^h}{\bm{r}} 
    + \sum_n \frac{ \braket[3]{\bm{r}'}{V^t}{\phi_{n}} \braket[3]{\phi_n} {V^t}{\bm{r} }}
    {E-E_n} , \label{eq:VDD}
\end{align}
where $\ket{\phi_n}$ and $E_n$ are the eigenstates and eigenenergies of the quark channel Hamiltonian, 
\begin{align}
    (T^q + V^q)\ket{\phi_n} = E_n\ket{\phi_n} ,
\end{align}
representing compact states composed purely of quark degrees of freedom. Owing to the confinement, there are only discrete eigenstates in this channel. $\ket{\bm{r}}$ is the eigenstate of the position operator $\bm{r}$ in the hadron channel.\footnote{In this paper, we adopt the convention of the position eigenstate $\ket{\bm{r}}$ and the momentum eigenstate $\ket{\bm{k}}$ as $\braket[2]{\bm{r}'}{\bm{r}}=\delta(\bm{r}'-\bm{r})$, $\braket[2]{\bm{k}'}{\bm{k}}=(2\pi)^3\delta(\bm{k}'-\bm{k})$, and $\braket[2]{\bm{r}}{\bm{k}}=e^{i\bm{k}\cdot\bm{r}}$.} It can be seen from the second term that the effective potential is nonlocal and energy dependent, as a consequence of the channel elimination.

Near the threshold of the hadron channel, which is of our interest in this study, we assume that the effective hadron interaction is dominated by the contribution from the $n=0$ state $\phi_0$ and contributions from other states can be neglected. We adopt the Yukawa-type form factor for the transition matrix element between the quark and hadron channels as
\begin{align}
    \braket[3]{\phi_0} {V^t}{\bm{r} } = g_0 V(\bm{r}), \quad V(\bm{r})=\frac{e^{-\mu r}}{r}, 
\end{align}
with $g_0$ being the coupling constant between the quark and hadron degrees of freedom. $\mu$ is the momentum cutoff of the form factor, as it can be seen from the momentum representation:
\begin{align}
       \Tilde{V}(k)
       &= \int d^3\bm{r}\ e^{-i \bm{k}\cdot\bm{r}} V(\bm{r})=\frac{4\pi}{\mu^2+k^2} .
       \label{eq:tildeVk}
\end{align}
Namely, this corresponds to the Yamaguchi-type form factor~\cite{Yamaguchi:1954mp,Yamaguchi:1954zz}. The direct hadron interaction is assumed to be a separable form with the same form factor as Eq.~\eqref{eq:tildeVk},
\begin{align}
     \braket[3]{\bm{r}'}{V^{h}}{\bm{r}} =\omega^h V(\bm{r}')V(\bm{r}),\label{eq:nonlocalV}
\end{align}
where $\omega^h$ is the coupling strength of the interaction. From this definition, we find that $\omega_{h}$ has mass dimension $+2$. Under these assumptions, the effective potential takes the form
\begin{align}\label{eq:VDDeff_omega_h}
    V(\boldsymbol{r'},\boldsymbol{r},E)   &= 
    \omega(E) \frac{e^{-\mu r'}}{r'}\frac{e^{-\mu r}}{r} , \\
    \omega(E)&= \omega^h+\omega^q(E), \label{eq:omegaE}\\
    \omega^q(E) &=\frac{g_0^2}{E-E_0},
\end{align}
where $\omega^h$ is the direct interaction in the hadron channel and $\omega^q(E)$ is the interaction generated by the coupling to the quark channel. 

By solving the effective single-channel Schr\"odinger equation for $\psi_{E} (\bm{r})=\braket[2]{\bm{r}}{h}$ at energy $E$:
\begin{align}
       -\frac{1}{2m}\nabla^2\psi_{E} (\bm{r})
       +\int d^3\bm{r}^\prime 
       V(\boldsymbol{r},\boldsymbol{r}',E)\psi_{E} (\bm{r}^\prime) &= E\psi_{E} (\bm{r}),
\end{align}
we obtain the bound ($E<0$) and scattering ($E>0$) solutions. Thanks to the separable form of the potential, the wave functions can be obtained analytically. The explicit form of the bound-state wave function with binding energy $E = -B$ is given by~\cite{Aoki:2021ahj,Terashima:2023tun}
\begin{align}\label{eq:psi_yukawa}
    \psi_{-B} (r)  = N_b \left( -\frac{\kappa e^{-\kappa r}}{r} + \frac{\kappa e^{- \mu r }}{r} \right),
\end{align}
where $\kappa = \sqrt{2m B}$. Here, $N_b$ is the normalization constant, which will be determined in the next section when we discuss the normalization of the wave function of the bound state by the energy-dependent potentials. It is worth noting that the wave function has a cutoff-dependent tail $e^{-\mu r}/r$ in addition to the one determined by the binding energy $e^{-\kappa r}/r$. This $\mu$-dependent tail reflects the nonlocal nature of the interaction. Furthermore, since this model has an artificial pole at $\kappa = \mu$ [see Eq.~\eqref{eq:TvG} below], deeply bound states with $\kappa > \mu$ are considered unphysical.

\subsection{$T$ matrix}

For the convenience of the later discission on the compositeness, we now solve the same problem using the momentum representation. In this case, the model possesses the same properties as the one proposed in Ref.~\cite{Takizawa:2012hy}. The on-shell $T$ matrix $T(E)$ is obtained from the Lippmann-Schwinger equation as~\cite{Kinugawa:2024crb}
\begin{align}
    T(E) = \frac{1}{\frac{1}{v(E)} - G(E)}\left(\frac{4\pi}{\mu^2+2mE}\right)^2,
    \label{eq:TvG}
\end{align}
where the interaction $v(E)$ and the loop function $G(E)$ are given by
\begin{align}
    v(E) &= \omega(E), \\
    G(E) &= \int \frac{d^3\bm{k}}{(2\pi)^3} \frac{1}{E - \frac{k^2}{2m} + i\epsilon} \left( \Tilde{V}(k) \right)^2.
\end{align}
With the form factor [Eq.~\eqref{eq:tildeVk}], the integration of the loop function is also analytically performed as
\begin{align}
    G(E) &= -\frac{4\pi m}{\mu\left(\sqrt{-2mE-i\epsilon}+\mu\right)^2} \label{eq:I_sekibun} .
\end{align}
If a bound state exists, the $T$ matrix has a pole at $E = -B$. Therefore, the condition for the existence of a bound state is given by
\begin{align}
        \frac{1}{v(-B)} - G(-B)  = 0, \label{eq:pole_cndtn}
\end{align}
and the binding energy $B$ can be determined by solving this equation. We note that Eq.~\eqref{eq:TvG} also possesses a pole at $E=-B_{b}$ with 
\begin{align}
    B_b = \frac{\mu^2}{2m} . 
    \label{eq:Bb}
\end{align}
This pole arises from the form factor and exists independently of the interaction; it does not represent a physical bound state but rather an artificial pole. Therefore, the present model cannot be extended to the region $E \leq -B_b$, and the physical bound states determined by the condition in Eq.~\eqref{eq:pole_cndtn} must exist within the range $0 \leq B \leq B_b$. In fact, $B_{b}$ corresponds to the typical energy scale determined by the cutoff $\mu$, and it sets the ultraviolet scale where the effective description breaks down in the perspective of the effective field theory~\cite{Kinugawa:2023fbf}. 

When a bound state exists at $E = -B$, the potential strength of the hadronic degrees of freedom, $\omega^h$, is bounded from below by the Hermiticity of the Hamiltonian, $g_0^2 \geq 0$. To see this, we first note that the loop function at $E = -B$ is negative:
\begin{align}\label{eq_Gnetative}
    G(-B) = -\frac{4\pi m}{\mu(\kappa + \mu)^2} < 0,
\end{align}
which essentially corresponds to the fact that the second-order perturbative correction to the ground-state energy is negative~\cite{Hyodo:2008xr}. From bound-state condition~\eqref{eq:pole_cndtn}, the coupling constant is expressed as
\begin{align}
    g_0^2 = (B + E_0)\left( -\frac{1}{G(-B)} + \omega^h \right).
    \label{eq:g02}
\end{align}
In the case of $\omega^h = 0$, the condition $g_0^2 \geq 0$ for a Hermitian Hamiltonian requires that $E_0 > -B$. This condition is natural, because a bare state with $E_0 < -B < 0$ gains only negative self-energy from the coupling to the scattering states above, preventing the pole from reaching $E = -B$. Under the condition $E_0 > -B$, we always have $B + E_0 < 0$, and therefore, the condition to have positive $g_0^2$ becomes\footnote{In principle, one can satisfy $g_0^2 > 0$ by simultaneously taking $E_0 < -B$ and $\omega^h < \omega^h_b$, but such a scenario requires an unnatural cancellation in the interaction strength and is therefore not considered here~\cite{Kinugawa:2023fbf}.}
\begin{align}\label{eq:omega_h_lim}
    \omega^h \geq \omega^h_b, \quad \omega^h_b = \frac{1}{G(-B)}.
\end{align}
From Eq.~\eqref{eq_Gnetative}, we also see that the lower bound is negative, $\omega^h_b < 0$. In the case of $\omega^h = \omega^h_b$, the coupling constant vanishes [$g_0^2 = 0$ from Eq.~\eqref{eq:g02}], which implies that there is no contribution from the quark degrees of freedom and the bound state is formed purely by the hadronic interaction. In other words, the bound state in this case is expected to be a purely hadronic molecule, which will be examined in the next section using the compositeness.

\subsection{Compositeness}

In this section, we analytically derive expressions for the compositeness, which serves as a quantitative measure of the internal structure of hadrons. The compositeness $X$ and the elementarity $Z$ are defined as the probabilities of finding the scattering states $\ket{\bm{p}}$ and the bare state $\ket{\phi_0}$, respectively, within the bound state $\ket{B}$~\cite{Hyodo:2013nka,Oller:2017alp,vanKolck:2022lqz,Kinugawa:2024crb}:
\begin{align}
    X = \int \frac{d^3\bm{p}}{(2\pi)^3}|\braket[2]{\bm{p}}{B}|^2,
    \quad
    Z = |\braket[2]{\phi_0}{B}|^2,
    \label{eq:XZdef}
\end{align}
where $\ket{\bm{p}}$ denotes the free scattering state in the hadron channel with momentum $\bm{p}$, which is an eigenstate of $T^h$:
\begin{align}
    T^h\ket{\bm{p}}
    =\frac{p^2}{2m}\ket{\bm{p}} .
\end{align}
It follows from the normalization of the bound state $\braket[2]{B}{B}=1$ and the completeness relation that
\begin{align}
    X+Z&=1 , \label{eq:normalization}
\end{align}
which ensures the normalization of the probabilities~\cite{Hyodo:2013nka,Kinugawa:2024crb}. There are two approaches to derive the explicit form of the compositeness: one based on the normalization of the wave function~\cite{Miyahara:2018onh}, and the other based on the Lippmann-Schwinger equation~\cite{Kamiya:2015aea,Kamiya:2016oao}. After deriving the compositeness using each method, we confirm that the obtained results are consistent with each other.

First, we derive the compositeness from the bound-state wave function. The normalization condition for a bound state with an energy-dependent and nonlocal potential is obtained from the requirement of the continuity equation as~\cite{CJP54.289,Miyahara:2015bya,Miyahara:2018onh}
\begin{align}\label{eq:X1_bnd_cnd}
    1 = \left. \int d^3\bm{r}' d^3\bm{r}  \psi^*_{E}(\bm{r}') 
    \left[
      \delta(\bm{r}' -\bm{r})-  \frac{\partial V(\bm{r}',\bm{r},E)}{\partial E} 
    \right]
    \psi_E(\bm{r}) \right|_{E=-B} .
\end{align}
By substituting the explicit forms of the potential [Eq.~\eqref{eq:VDDeff_omega_h}] and the wave function [Eq.~\eqref{eq:psi_yukawa}] into the normalization condition, we obtain
\begin{align}
    1 &= N_b^2(x+z) , \\
    x &=\int d^3\bm{r} \left( -\frac{\kappa e^{-\kappa r}}{r} + \frac{\kappa e^{- \mu r }}{r} \right)^2  \nonumber \\
    &= \frac{2\pi(\mu-\kappa)^2}{\mu\kappa(\mu+\kappa)}, \\
    z &= -\int d^3\bm{r}' d^3\bm{r}
    \left( -\frac{\kappa e^{-\kappa r'}}{r'} + \frac{\kappa e^{- \mu r' }}{r'} \right)
    \frac{e^{- \mu r' }}{r'}\nonumber \\
    &\quad \times \left(-\frac{g_0^2}{(-B-E_0)^2}\right)
    \frac{e^{- \mu r }}{r}
    \left( -\frac{\kappa e^{-\kappa r}}{r} + \frac{\kappa e^{- \mu r }}{r} \right)
    \nonumber \\
    &= \frac{g_0^2}{(B+E_0)^2}\frac{(4\pi)^2(\mu-\kappa)^2}{4\mu^2(\mu+\kappa)^2}.
\end{align}
Thus, the normalization constant is determined as
\begin{align}
    N_b
    &=\sqrt{\frac{\mu\kappa(\mu+\kappa)}{2\pi(\mu-\kappa)^2}}
    \left[1 +   \frac{g_0^2}{(B+E_0)^2}\frac{2\pi\kappa}{\mu(\mu +\kappa)} \right]^{-1/2}
    .
\end{align}
In the present framework, the definition of the compositeness $X$ and the elementarity $Z$ in Eq.~\eqref{eq:XZdef} can be expressed as~\cite{Miyahara:2018onh}
\begin{align}
    X
    &=\int d^3\bm{r} |\psi_{-B}(\bm{r}) |^2=N_b^2x ,
    \label{eq:compositeness} \\
    Z
    &=N_b^2z
    =1-X.
\end{align}
From the above result, we obtain the analytic expression of the compositeness
\begin{align}
  X &=  \left[1 +   \frac{g_0^2}{(B+E_0)^2}\frac{2\pi\kappa}{\mu(\mu +\kappa)} \right]^{-1}. \label{eq:X_bound}
\end{align}
The bound-state wave function can then be expressed by using the compositeness as
\begin{align}
     \psi_{-B} (r)=\sqrt{X}\sqrt{\frac{\mu\kappa(\mu+\kappa)}{2\pi(\mu-\kappa)^2}}
    \left(\frac{e^{-\mu r}}{r}-\frac{e^{-\kappa r}}{r}\right) .
    \label{eq:BSwf}
\end{align}
Since this wave function represents the scattering component of the hadron channel, it is clear that the magnitude of the wave function decreases as the compositeness $X$ becomes smaller.

Next, we derive the compositeness using the Lippmann-Schwinger equation. When the on-shell scattering amplitude can be expressed in terms of $v(E)$ and $G(E)$ as in Eq.~\eqref{eq:TvG}, the compositeness and  elementarity are calculated as~\cite{Kamiya:2015aea,Kamiya:2016oao}
\begin{align}
    X&=\left.\frac{-G'(E)}{[v^{-1}(E)]' -G'(E)} \right|_{E=-B}, \\
    Z&=\left.\frac{[v^{-1}(E)]'}{[v^{-1}(E)]'-G'(E)} \right|_{E=-B}.
\end{align}
Substituting the explicit forms given in Eqs.~\eqref{eq:omegaE} and \eqref{eq:I_sekibun}, we obtain
\begin{align}
X &=  \bab{1+\frac{g_0^2\kappa\mu(\kappa+\mu)^3}{8\pi m^2(g_0^2-(B+E_0)\omega^h)^2}}^{-1} 
\end{align}
Although this result appears different from Eq.~\eqref{eq:X_bound}, they must agree because both are based on the same definition given in Eq.~\eqref{eq:XZdef}.  
Indeed, by using bound-state condition~\eqref{eq:pole_cndtn},
\begin{align}
    G(-B)\omega(-B) = 1,
\end{align}
to eliminate $\omega_h$, one can analytically show that the two expressions are identical.

\subsection{Scattering observables}

In this model, the $T$ matrix is analytically obtained as shown in Eq.~\eqref{eq:TvG}, which allows for analytical expressions of various scattering observables. The scattering amplitude $f(k)$ is given by~\cite{Kinugawa:2024crb}
\begin{align}
    f(k) = -\frac{m}{2\pi}T(E) ,
    \label{eq:fk}
\end{align}
with $k = \sqrt{2mE}$. From the scattering amplitude, the phase shift $\delta(k)$ can be defined through
\begin{align}
    f(k) = \frac{1}{k\cot\delta(k) - ik}.
\end{align}
Substituting Eq.~\eqref{eq:TvG} into Eq.~\eqref{eq:fk}, we obtain the explicit expression
\begin{align}
    k \cot\delta(k)
     & = -\frac{1}{8\pi m}
     \left[
     \frac{(\mu^{2}+k^{2})^{2}}{\omega(E)}
     +\frac{4\pi m(\mu^{2}-k^{2})}{\mu}
     \right],
     \label{eq:kcotdelta}
\end{align}
which reproduces Eq.~(48) of Ref.~\cite{Terashima:2023tun}. 

By substituting $\omega(E)$ in Eq.~\eqref{eq:omegaE}, and comparing with the effective range expansion,
\begin{align}
    k \cot\delta(k)
    = -\frac{1}{a_{0}}+\frac{r_{e}}{2}k^{2}+\mathcal{O}(k^{4}),
\end{align}
we obtain the scattering length $a_0$ and the effective range $r_e$ as
\begin{align}
    a_{0}
    &= 
    \frac{1}{\mu}
    \left[
    \frac{8\pi m }{4\pi m+\mu^3/(\omega^h-g_0^2/E_0)}
    \right],
    \label{eq:a0}\\
    r_e
    &=\frac{1}{\mu}-\frac{\mu^2}{2\pi m(\omega^h-g_0^2/E_0)}-\frac{\mu^4g_0^2}{8\pi m^2E_0^2(\omega^h-g_0^2/E_0)^2}.
    \label{eq:re}
\end{align}
Here, the first term of the effective range, $1/\mu > 0$, represents the typical length scale of the interaction determined by the form factor, while the second and third terms provide corrections to this reference scale $1/\mu$. The quantity $\omega^{h} - g_{0}^{2}/E_{0}$ corresponds to the value of the interaction $v(E)$ at the threshold $E = 0$, and it is negative (i.e., attractive) in the absence of fine tuning. Therefore, the second term in Eq.~\eqref{eq:re} gives a positive contribution, while the third term gives a negative one to $1/\mu$.

\subsection{Local approximation}\label{subsec:local}

Although the potential in Eq.~\eqref{eq:nonlocalV} is nonlocal, for practical applications such as few-body calculations, a local approximation of the potential is often more useful. By substituting $V(\bm{r}^{\prime},\bm{r},E)=V(\bm{r},E)\delta(\bm{r}^{\prime}-\bm{r})$ into the normalization condition in Eq.~\eqref{eq:X1_bnd_cnd},  
the normalization condition for a local energy-dependent potential $V(\bm{r},E)$ is given by
\begin{align}
        1 = \left. \int d^{3}\bm{r}  \psi^*_{E}(\bm{r}) 
    \left[
      1-  \frac{\partial  V(\bm{r},E)}{\partial E}
    \right]
    \psi_E(\bm{r}) \right|_{E=-B},
    \label{eq:bclocal}
\end{align}
as shown in Refs.~\cite{CJP54.289,Miyahara:2015bya}.  
Just as in the nonlocal case, it is evident that the energy dependence of the potential modifies the normalization condition.

There are several methods for the local approximation of a nonlocal potential. In this study, we adopt the derivative expansion by the HAL QCD method, which has been shown to be effective in previous studies~\cite{Aoki:2021ahj,Terashima:2023tun}. The HAL QCD method provides a systematic approach to constructing a local potential that reproduces the scattering wave functions at given momenta $k_i\ (i=0,1,\dots,n)$ obtained from the nonlocal potential. The accuracy of the approximation improves as the order of the derivative expansion increases. Since the scattering wave functions are reproduced, the phase shifts at each momentum $k_i$ are exactly reproduced.

As pointed out in Ref.~\cite{Terashima:2023tun}, the local potential approximated by the HAL QCD method depends on the choice of $k_i$ used to define it, but it does not possess explicit energy dependence. That is, the local potential by the HAL QCD method satisfies
\begin{align}
     \frac{\partial  V(\bm{r},E)}{\partial E} = 0 \quad (\text{HAL QCD method}) ,
\end{align}
and thus, from Eqs.~\eqref{eq:bclocal} and~\eqref{eq:compositeness}, the compositeness is given by
\begin{align}
      X = 1 \quad (\text{HAL QCD method}) ,
      \label{eq:XHAL}
\end{align}
indicating that it is always unity.

In this study, we construct a local potential using the leading order ($n=0$) derivative expansion with the momentum chosen at the threshold ($k_0 = 0$), which has been shown in Refs.~\cite{Aoki:2021ahj,Terashima:2023tun} to accurately describe the scattering observables near the threshold. In this case, since the wave function and the phase shift at $k = 0$ coincide with those by the nonlocal one, the scattering length reproduces Eq.~\eqref{eq:a0}. However, the phase shifts at finite momenta must be obtained numerically by solving the Schr\"odinger equation. The effective range can also be determined numerically from the relation:
\begin{align}
       r_e= 2\left.
       \frac{d [k\cot{\delta(k)}]}{dk^2} 
       \right|_{k\to0}
       \quad (\text{HAL QCD method}) .
\end{align}

\section{Numerical results}\label{sec:results}

\subsection{Reference model}

In this section, we take the $X(3872)$, a typical weakly bound state, as an example and investigate the relationship between the compositeness and observable quantities such as the scattering phase shift. First, we construct a model that reproduces the $X(3872)$ using the parameter values shown in Table~\ref{tab:param_X3872}, following Ref.~\cite{Terashima:2023tun}. The binding energy $B$ is taken from the central value reported by the Particle Data Group (PDG)~\cite{ParticleDataGroup:2024cfk}. The bare energy $E_0$ of the $c\bar{c}$ state is adopted from the constituent quark model prediction~\cite{Godfrey:1985xj} for the $\chi_{c1}(2P)$ mass. The cutoff $\mu$ is chosen to correspond to the pion mass, as estimated by the one-pion exchange picture. The strength of the direct interaction $\omega^h$ is taken to be $\omega^h = 0$ as a reference value, as in Ref.~\cite{Terashima:2023tun}. The coupling constant $g_0$ is determined so that the bound-state condition in Eq.~\eqref{eq:pole_cndtn} is satisfied for the given binding energy $B$ and the other parameters. 

\begin{table}[tbp]
    \centering
    \caption{Binding energy and parameters of the reference model.}
\begin{ruledtabular}
    \begin{tabular}{clrc}
       & Parameter                     & Reference value &\\ \hline
       & Binding energy $B$            & $0.04$ MeV &\\
       & Bare energy $E_0$             & $78$ MeV & \\
       & Cutoff $\mu$                  & $140$ MeV &\\
       & Direct interaction $\omega^h$ & 0 MeV$^{2}$ &\\
    \end{tabular}
\end{ruledtabular}
    \label{tab:param_X3872}
\end{table}

With this reference model, the compositeness is calculated as
\begin{align}
       X = 0.99, \label{eq:Xref}
\end{align}
indicating that the state is predominantly composite due to its weak binding. The scattering length and effective range are found to be
\begin{align}
       a_0 = 24.5 \ {\rm fm}, \quad r_e = 3.74 \ {\rm fm}.
       \label{eq:a0reref}
\end{align}
The effective range is of the order of the typical length scale $1/\mu \approx 1.41$ fm, while the much larger scattering length is also a characteristic feature of weakly bound states, reflecting the consequence of the low-energy universality~\cite{Braaten:2004rn,Naidon:2016dpf,Kinugawa:2022fzn}.

Based on the above setup, we investigate how the compositeness and observable quantities change by varying the binding energy $B$ or other model parameters from the values listed in Table~\ref{tab:param_X3872}. When varying $B$, we fix $E_0$, $\mu$, and $\omega^h$ to the values in Table~\ref{tab:param_X3872}, and adjust the coupling constant $g_0$ so as to reproduce the new value of $B$. When varying one of the parameters $E_0$, $\mu$, or $\omega^h$, we fix the remaining parameters to their values in Table~\ref{tab:param_X3872}, and tune $g_0$ such that the binding energy $B$ remains the same as in Table~\ref{tab:param_X3872}.

\subsection{Variation of binding energy $B$}

\begin{figure}[tbp]
    \centering
    \includegraphics[width=1\linewidth]{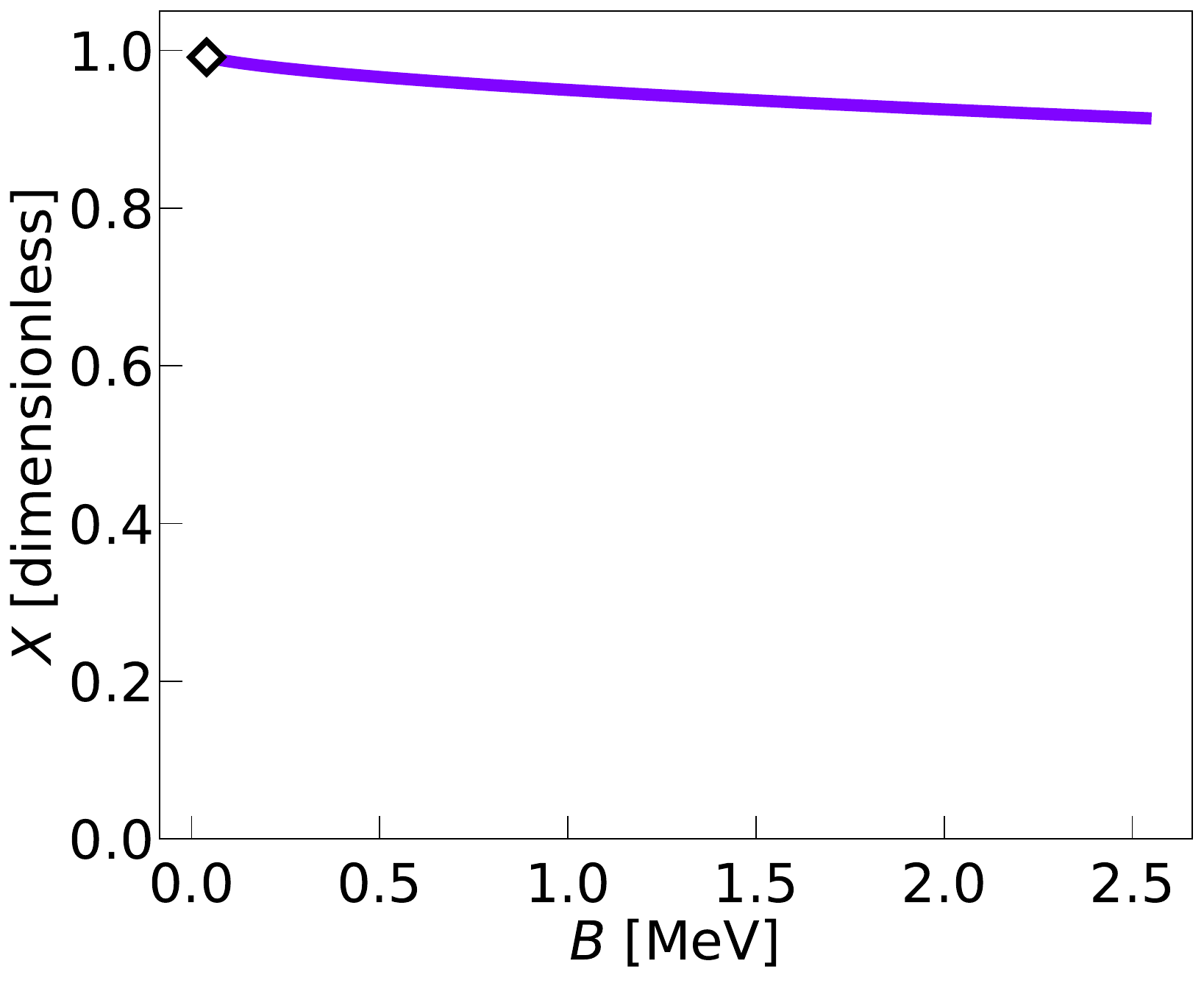}
    \caption{Compositeness $X$ as a function of the binding energy $B$. Reference value in Table~\ref{tab:param_X3872} is indicated by the diamond.}
    \label{fig:X_B}
\end{figure}

Figure~\ref{fig:X_B} shows the compositeness $X$ as a function of the binding energy $B$. Starting from the weakly bound state characterized by the reference value in Table~\ref{tab:param_X3872}, we vary $B$ up to $B_b/4 \simeq 2.53$~MeV, where $B_b$ is the upper limit of the binding energy given by Eq.~\eqref{eq:Bb}. The result at the binding energy in Table~\ref{tab:param_X3872} is marked with a diamond in the figure. As shown in the figure, the compositeness $X$ decreases as the binding energy $B$ increases. This behavior is consistent with the known result that $X$ approaches 1 as $B$ becomes small~\cite{Sazdjian:2022kaf,Lebed:2022vks,Kinugawa:2023fbf,Song:2023pdq}, and that $X \to 1$ in the strict limit $B \to 0$ \cite{Hyodo:2014bda}. It is also worth noting that, even at $B = B_b/4$, the compositeness remains as large as $X = 0.91$, indicating only a moderate deviation from the reference value given in Eq.~\eqref{eq:Xref}. As we will see later, this is due to the fact that the bare energy $E_0$ is sufficiently large, which suppresses the contribution from the elementary component to the bound state.

\begin{figure}[tbp]
    \centering
    \includegraphics[width=1\linewidth]{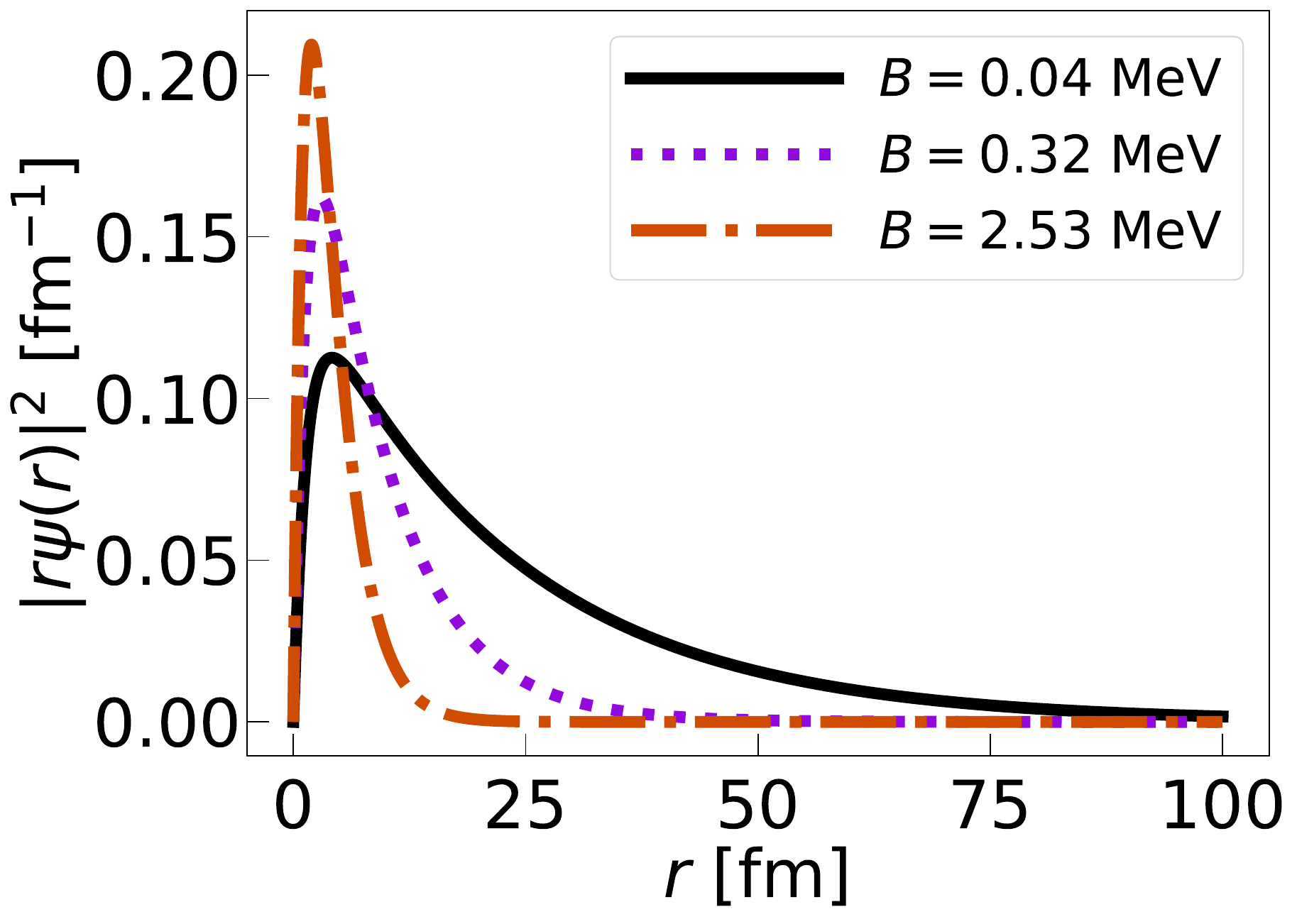}
    \caption{Bound-state wave functions $|r\psi(r)|^2$ for $B = 0.04$~MeV (reference value, solid line), $B = 0.32$~MeV (dashed line), and $B = B_b/4 \simeq 2.53$~MeV (dot-dashed line). }
    \label{fig:w-f_B}
\end{figure}

Figure~\ref{fig:w-f_B} shows the bound-state wave functions given by Eq.~\eqref{eq:BSwf} for $B = 0.04$~MeV (reference value, solid line), $B = 0.32$~MeV (dashed line), and $B = B_b/4 \simeq 2.53$~MeV (dot-dashed line). Since the wave function asymptotically behaves as $r\psi(r)\sim \exp\{-\sqrt{2mB} \, r\}$ for $r \to \infty$, the spatial extent of the wave function becomes narrower as the binding energy increases. According to Eq.~\eqref{eq:compositeness}, when there is a coupling to the elementary component, the volume integral of the squared scattering wave function becomes the compositeness $X$ instead of unity. Therefore, if $X$ deviates from unity, the norm of the wave function is reduced accordingly. As shown in Fig.~\ref{fig:X_B}, the compositeness does not significantly change even when $B$ is increased up to $B_b/4 \simeq 2.53$~MeV. Thus, as $B$ increases and the wave function becomes more localized near the origin, its amplitude also becomes larger, as seen in Fig.~\ref{fig:w-f_B}.

\begin{figure}[tbp]
    \centering
    \includegraphics[width=1\linewidth]{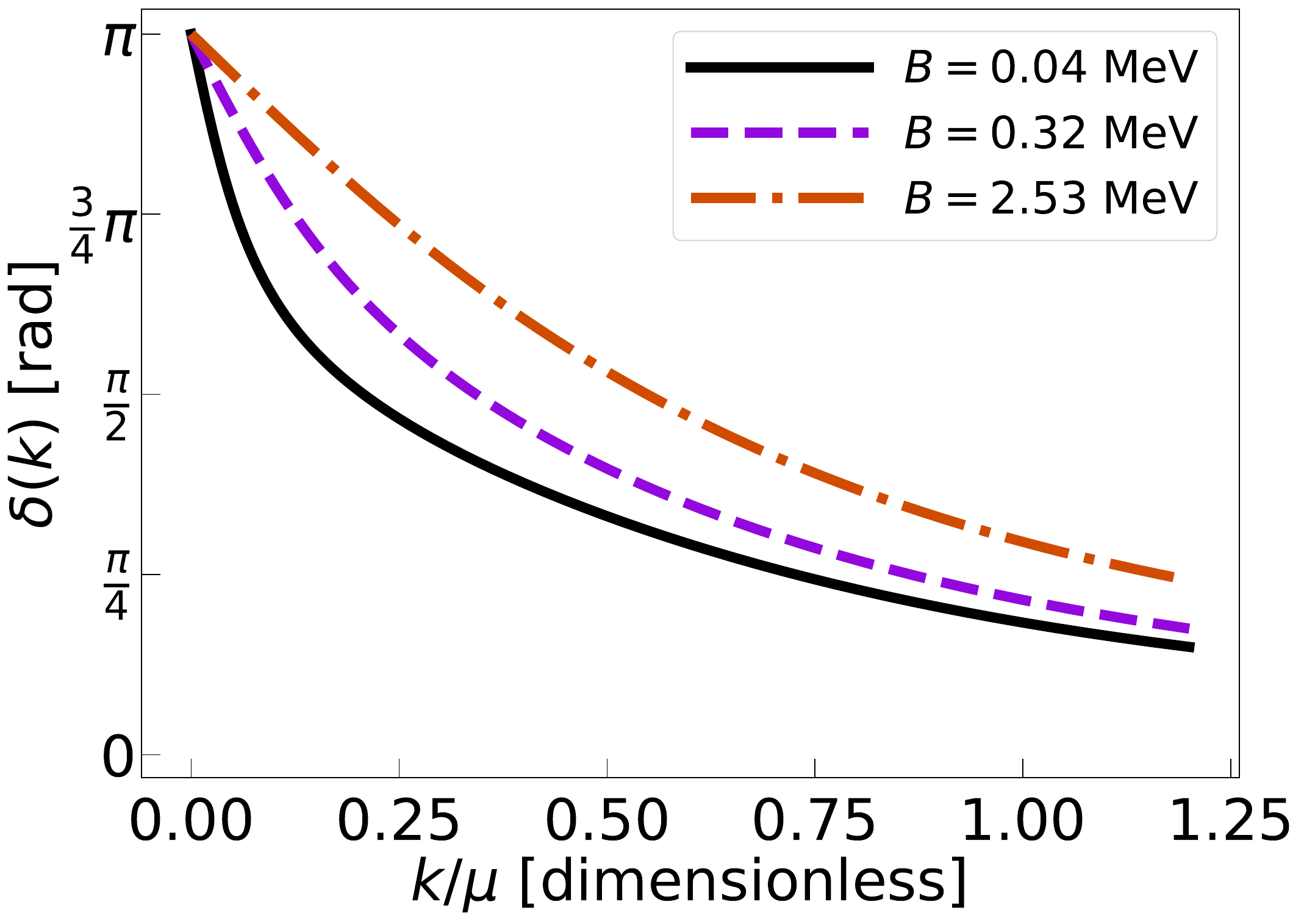}
    \caption{Phase shifts $\delta$ as functions of $k/\mu$ for $B = 0.04$~MeV (reference value, solid line), $B = 0.32$~MeV (dashed line), and $B = B_b/4 \simeq 2.53$~MeV (dot-dashed line).}
    \label{fig:delta_B}
\end{figure}

To investigate how the difference in the compositeness and wave function affects observable quantities, we show in Fig.~\ref{fig:delta_B} the phase shifts as functions of $k/\mu$ for three binding energies: $B = 0.04$~MeV (reference value, solid line), $B = 0.32$~MeV (dashed line), and $B = B_b/4 \simeq 2.53$~MeV (dot-dashed line). As $B$ increases, the slope of the phase shift near $k \approx 0$ decreases, indicating a weaker momentum dependence of the phase shift. This behavior can be understood as follows: the steep decrease in the phase shift at $k \approx 0$ in the $B = 0.04$~MeV case reflects the presence of a weakly bound state close to the threshold. As the binding energy increases and the bound-state pole moves away from the threshold, the influence of the bound state on the low-energy phase shift becomes milder, resulting in a more gradual variation of the phase shift.

\begin{figure}[tbp]
    \centering
    \includegraphics[width=1\linewidth]{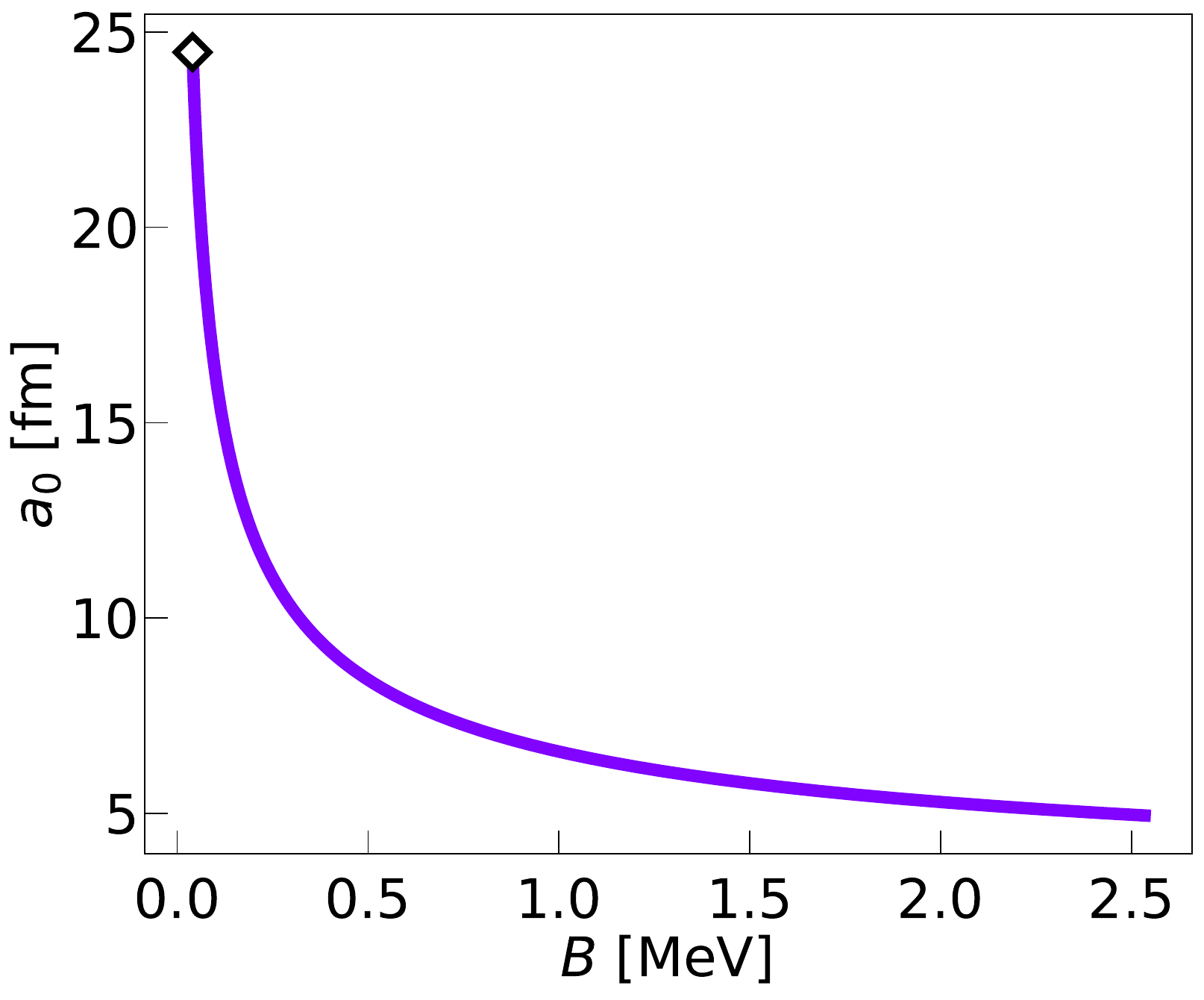}
    \caption{Scattering length $a_0$ as a function of the binding energy $B$. Reference value in Table~\ref{tab:param_X3872} is indicated by the diamond.}
    \label{fig:a0_B}
\end{figure}

\begin{figure}[tbp]
    \centering
    \includegraphics[width=1\linewidth]{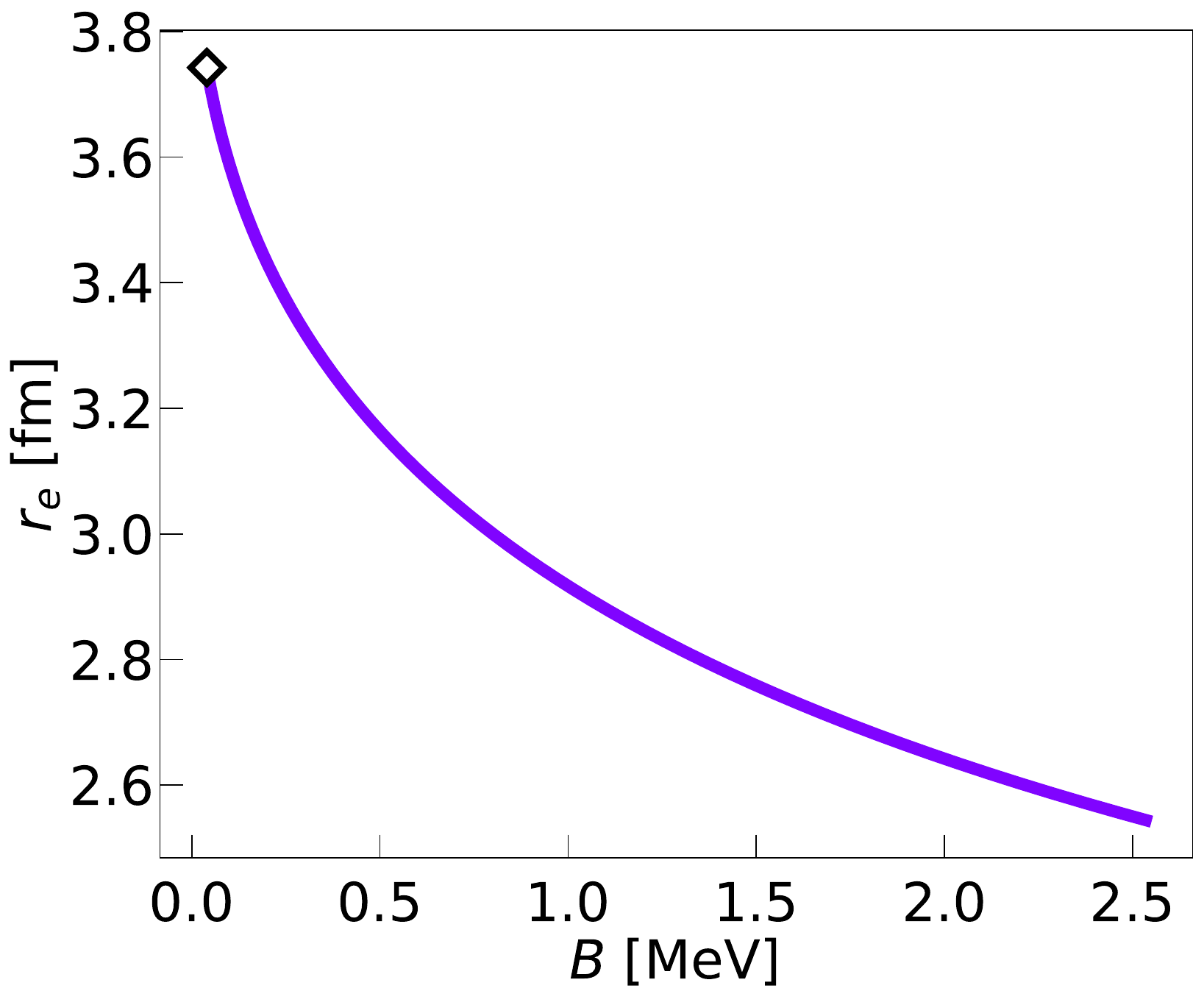}
    \caption{Effective range $r_e$ as a function of the binding energy $B$. Reference value in Table~\ref{tab:param_X3872} is indicated by the diamond.}
    \label{fig:re_B}
\end{figure}

To quantitatively characterize the variation of the phase shifts, we show the scattering length $a_{0}$ and effective range $r_{e}$ as functions of the binding energy $B$ in Figs.~\ref{fig:a0_B} and \ref{fig:re_B}, respectively. The scattering length $a_{0}$ rapidly decreases from the large value given by Eq.~\eqref{eq:a0reref} as $B$ increases, and asymptotically approaches approximately 5~fm around $B \simeq 2.53$~MeV. This behavior is consistent with the universal relation $a_{0} \sim 1/\sqrt{2 m B}$ when $X \approx 1$. Since the slope of the phase shift at $k \approx 0$ is determined by the scattering length, the rapid decrease in $a_{0}$ explains the behavior of the phase shift observed in Fig.~\ref{fig:delta_B}. The effective range $r_{e}$ also decreases as $B$ increases, but its variation in the plotted range is about 1.2~fm, showing a relatively smaller change compared to the scattering length. This is because the reference value of $r_{e}$ is 3.74~fm, which is not particularly large, so the degree of variation is relatively modest.

\subsection{Variation of bare energy $E_0$}
\label{subsec:variationE0}

\begin{figure}[tbp]
    \centering
    \includegraphics[width=1\linewidth]{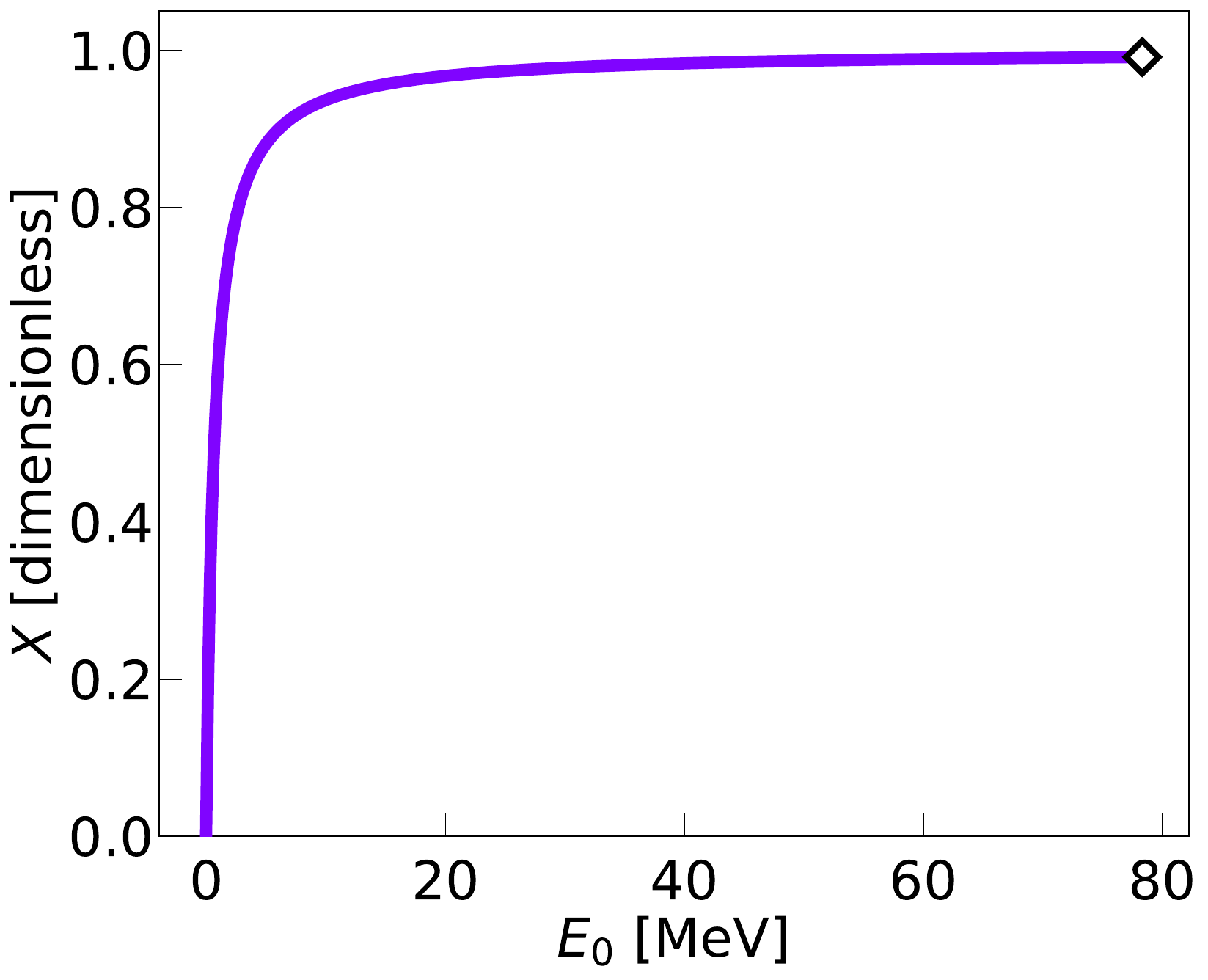}
    \caption{Compositeness $X$ as a function of the bare energy $E_{0}$. Reference value in Table~\ref{tab:param_X3872} is indicated by the diamond.}
    \label{fig:X_E0}
\end{figure}

Next, we investigate the relationship between the compositeness and the scattering observables when the bare energy $E_{0}$ is varied. Figure~\ref{fig:X_E0} shows the result of the compositeness as a function of $E_{0}$ when it is varied from the reference value down to $-B$. As $E_{0}$ decreases from the reference value, the compositeness gradually decreases, then rapidly drops in the vicinity of $E_{0} \sim -B$, and reaches $X=0$ at $E_{0} = -B$. This behavior can be interpreted as follows. The bare state located at $E = E_{0}$ is composed of the quark channel, and thus has a compositeness of $X = 0$. When $E_{0} = -B$, the bare state becomes the physical bound state without acquiring any loop correction which induces the hadronic molecular component, so the compositeness is $X = 0$. However, for $E_{0} > -B$, the bare state acquires a self-energy through the coupling with the hadronic scattering channel and evolves into the physical bound state at $E = -B$, which gives rise to a finite compositeness. The fact that the compositeness is $X \approx 1$ in most of the parameter range shown in Fig.~\ref{fig:X_E0} is a characteristic feature of weakly bound states with small binding energy $B$. It indicates that, in order to realize a shallow bound state with $X \approx 0$, one must fine-tune the bare energy $E_{0}$ to be within a narrow range near $E_{0} \sim -B$~\cite{Kinugawa:2023fbf}.

\begin{figure}[tbp]
    \centering
    \includegraphics[width=1\linewidth]{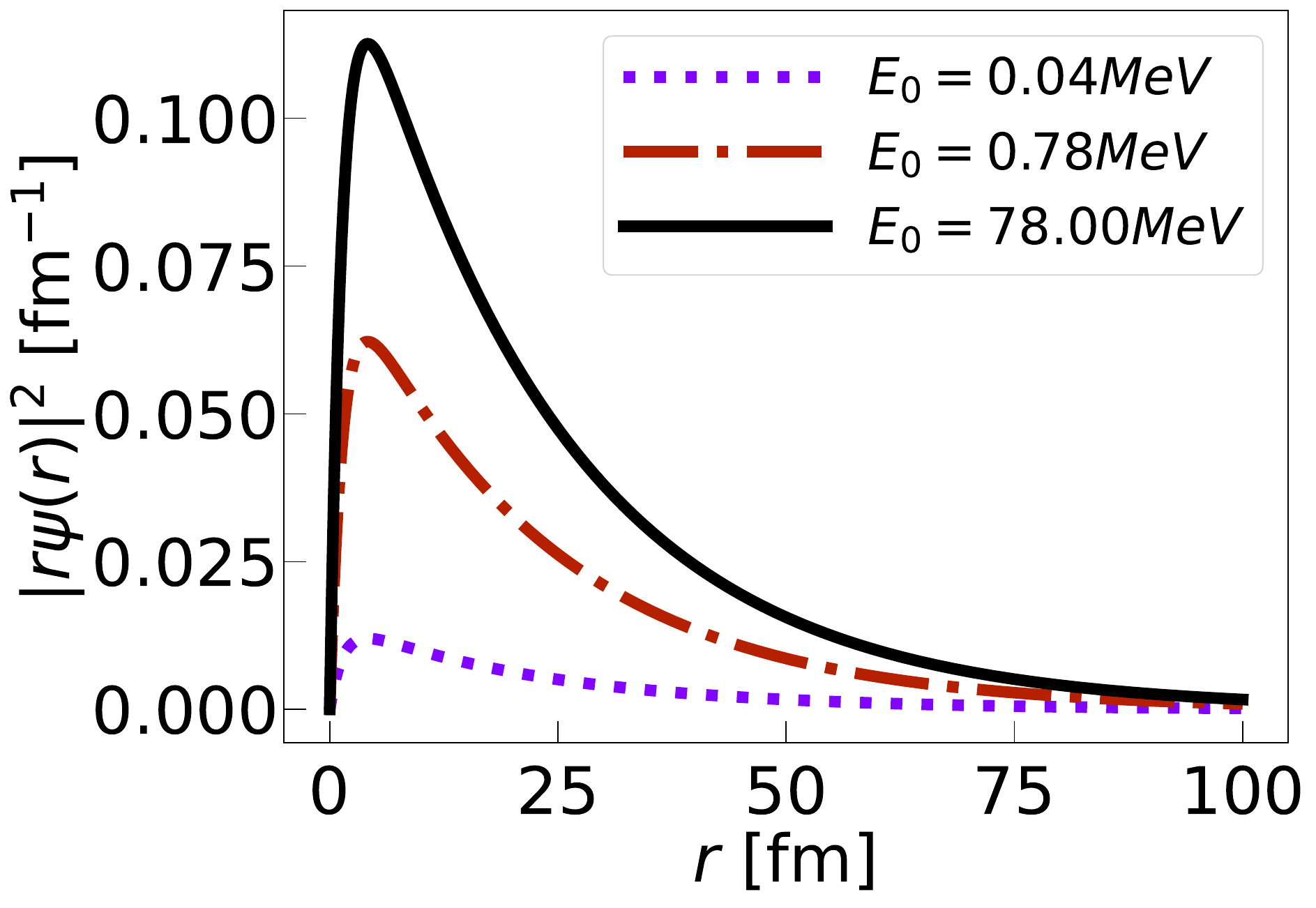}
    \caption{Bound-state wave functions $|r\psi(r)|^2$ for $E_0 = 78$~MeV (reference value, solid line), $E_0 = 0.78$~MeV (dashed line), and $E_0=0.04$~MeV (dot-dashed line).}
    \label{fig:w-f_E0}
\end{figure}

Figure~\ref{fig:w-f_E0} shows the bound-state wave functions for $E_{0} = 78$~MeV ($X = 0.99$), $E_{0} = 0.78$~MeV ($X = 0.55$), and $E_{0} = +B = 0.04$~MeV ($X = 0.11$). Unlike Fig.~\ref{fig:w-f_B}, the binding energy $B$ is fixed in this case, and therefore the asymptotic behavior of the wave functions at large $r$ remains unchanged. However, as $E_{0}$ decreases, the compositeness $X$ also becomes smaller, leading to a visible reduction in the magnitude of the wave function. This indicates that the change in compositeness manifests as a change in the norm of the bound-state wave function.

\begin{figure}[tbp]
    \centering
    \includegraphics[width=1\linewidth]{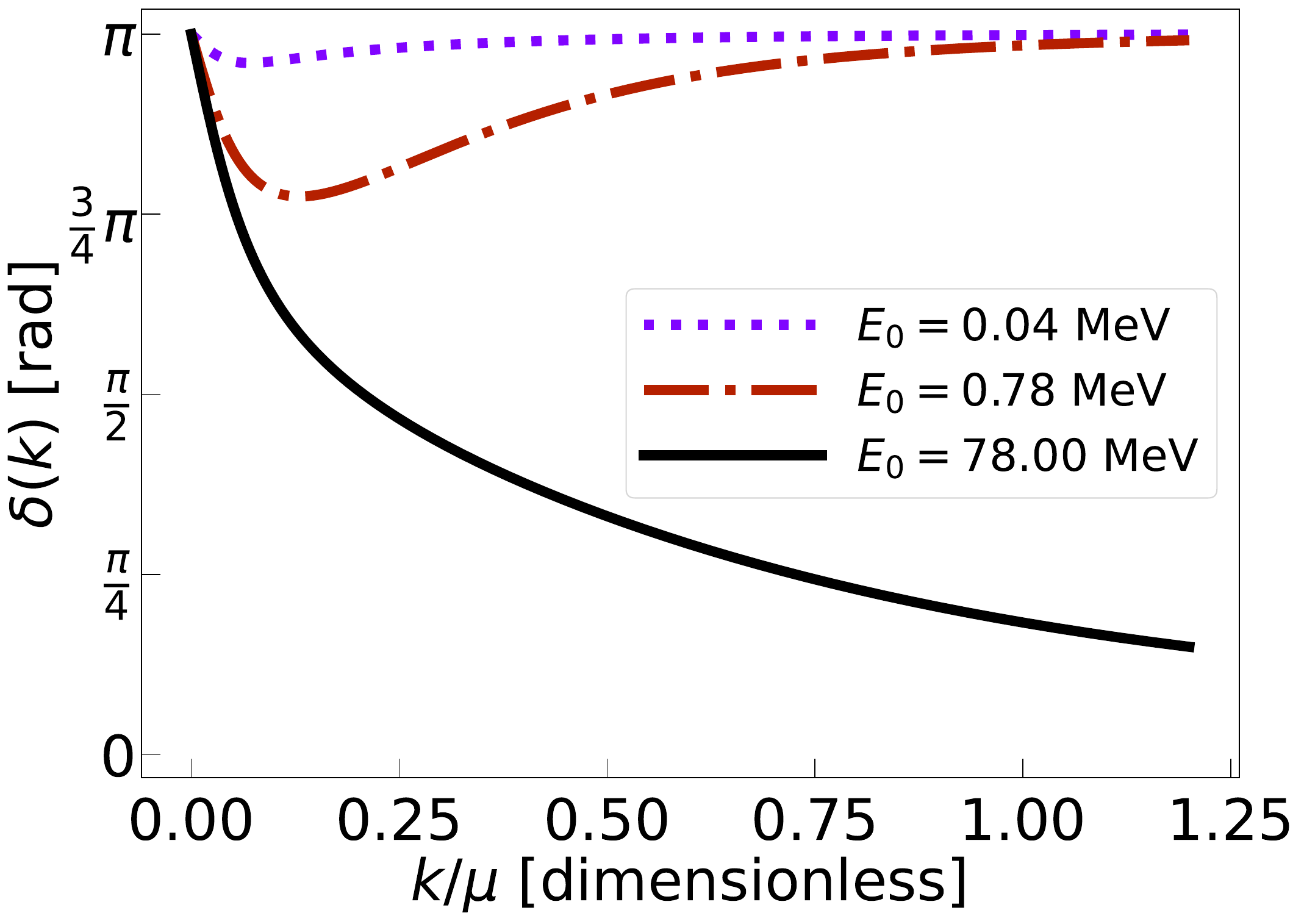}
    \caption{Phase shifts $\delta$ as functions of $k/\mu$ for $E_{0} = 78$~MeV (reference value, solid line), $E_{0} = 1.77$~MeV (dashed line), and $E_{0} = 0.04$~MeV (dot-dashed line).}
    \label{fig:delta_E0}
\end{figure}

Next, we calculate the phase shifts for the three choices of $E_{0}$, and plot the results in Fig.~\ref{fig:delta_E0} as functions of the dimensionless momentum $k/\mu$. As the value of $E_{0}$ decreases, the phase shift becomes closer to $\delta = \pi$. This tendency is similar to the case in Fig.~\ref{fig:delta_B}, where the binding energy $B$ is increased, but the variation of the phase shift is more pronounced when $E_{0}\sim -B$ is chosen. This stronger dependence can be understood from the comparison between Figs.~\ref{fig:X_B} and \ref{fig:X_E0}, which shows that the corresponding change in the compositeness $X$ is larger in the case of varying $E_{0}$. In particular, since the present calculation fixes the binding energy $B$ while changing the compositeness, Fig.~\ref{fig:delta_E0} reveals the effect of the internal structure of a weakly bound state on the scattering behavior above the threshold.

\begin{figure}[tbp]
    \centering
    \includegraphics[width=1\linewidth]{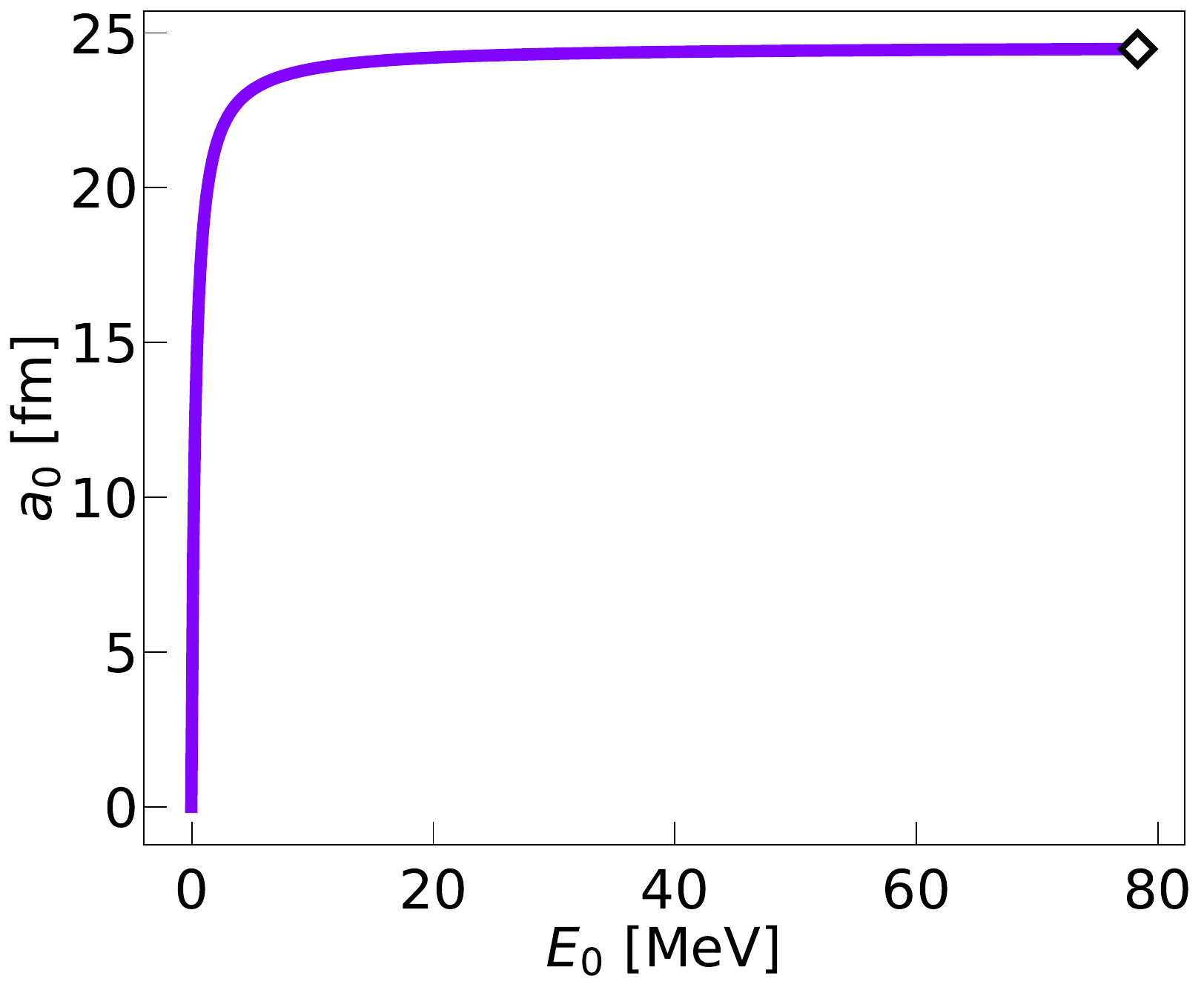}
    \caption{Scattering length $a_0$ as a function of the bare energy $E_{0}$. Reference value in Table~\ref{tab:param_X3872} is indicated by the diamond.}
    \label{fig:a0_E0}
\end{figure}

\begin{figure}[tbp]
    \centering
    \includegraphics[width=1\linewidth]{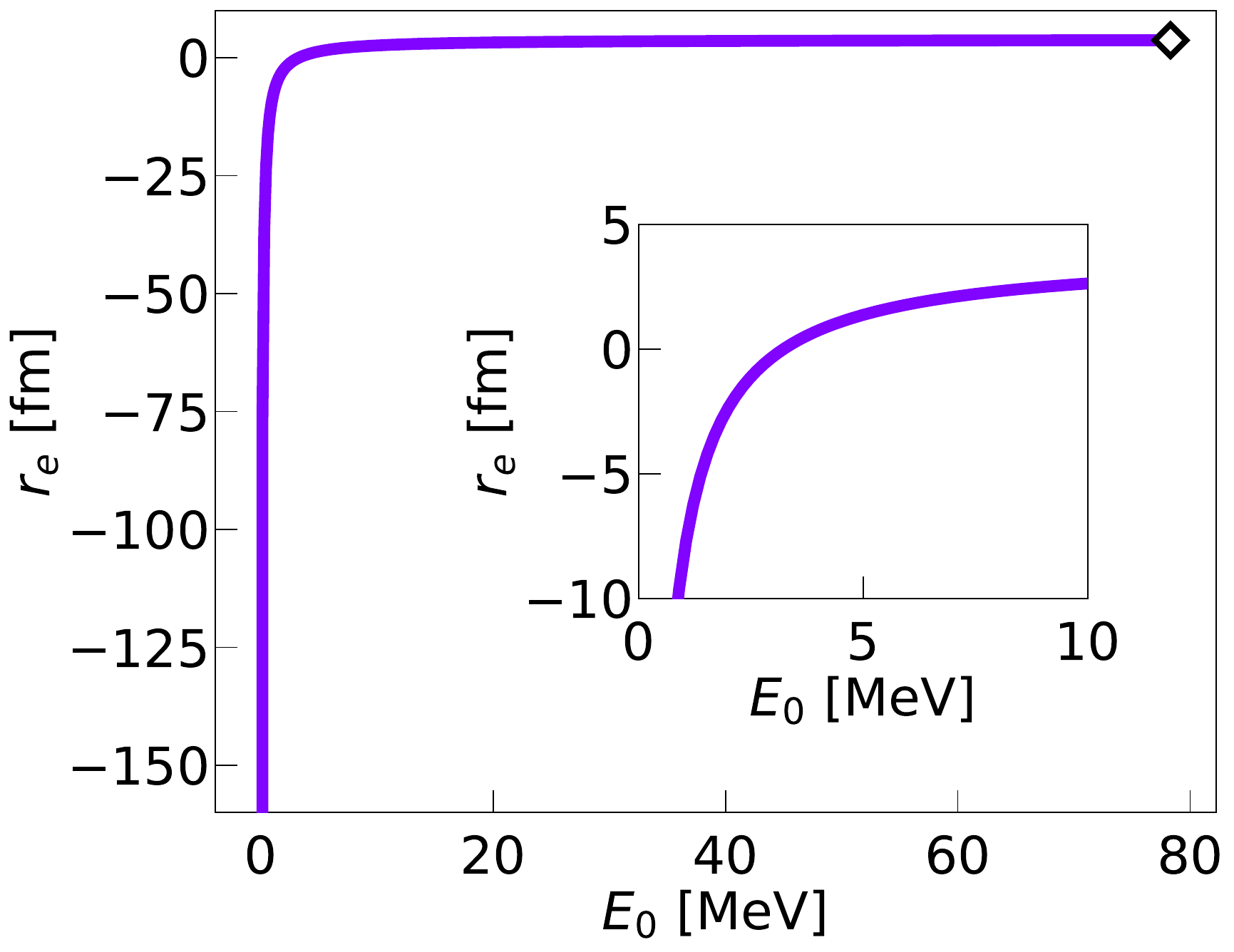}
    \caption{Effective range $r_e$ as a function of the bare energy $E_{0}$. Reference value in Table~\ref{tab:param_X3872} is indicated by the diamond. }
    \label{fig:re_E0}
\end{figure}

The dependence of the scattering length $a_0$ and the effective range $r_e$ on the bare energy $E_0$ is shown in Figs.~\ref{fig:a0_E0} and \ref{fig:re_E0}, respectively. It is evident that both $a_0$ and $r_e$ exhibit significant variations in the region where the compositeness $X$ changes rapidly, particularly near $X \approx 0$ around $E_{0}\to -B$. This is because the binding energy is sufficiently small, making the compositeness strongly correlated with the scattering parameters. In general, for a weakly bound state, the following relations are known to hold with $R = 1/\sqrt{2mB}$:
\begin{align}
   a_0 &= \frac{2X}{X+1} R + \mathcal{O}(R_{\rm typ}), 
   \label{eq:a0WBR}\\
   r_e &= \frac{X - 1}{X} R + \mathcal{O}(R_{\rm typ}),
   \label{eq:reWBR}
\end{align}
where $R_{\rm typ}$ denotes the typical length scale of the system~\cite{Weinberg:1965zz,Hyodo:2013nka,Kinugawa:2022fzn,Kinugawa:2024crb}. When $B$ is sufficiently small such that $R \gg R_{\rm typ}$, the $\mathcal{O}(R_{\rm typ})$ corrections can be neglected, and the scattering length and effective range are directly related to the compositeness $X$. In the present case, $R_{\rm typ} \sim 1/\mu \approx 1.4$ fm and $R \approx 22$ fm, so the condition for a weakly bound state is satisfied. Taking the limits $X \to 1$ and $X \to 0$ in Eqs.~\eqref{eq:a0WBR} and \eqref{eq:reWBR}, we find:
\begin{align}
   a_0 &\to R, \quad 
   r_e \to \mathcal{O}(R_{\rm typ}) \quad (X \to 1), \\
   a_0 &\to \mathcal{O}(R_{\rm typ}), \quad 
   r_e \to -\infty \quad (X \to 0).
\end{align}
Therefore, in Figs.~\ref{fig:a0_E0} and \ref{fig:re_E0}, except for the small-$E_{0}$ region, $a_{0}$ is comparable with $R \approx 22$ fm and $r_{e}$ with $R_{\rm typ} \approx  1.4$ fm, because $X\approx 1$. In the limit $E_0 \to -B$ where $X \to 0$, the scattering length approaches zero, and the effective range diverges negatively.

\subsection{Variation of cutoff $\mu$}

\begin{figure}[tbp]
    \centering
    \includegraphics[width=1\linewidth]{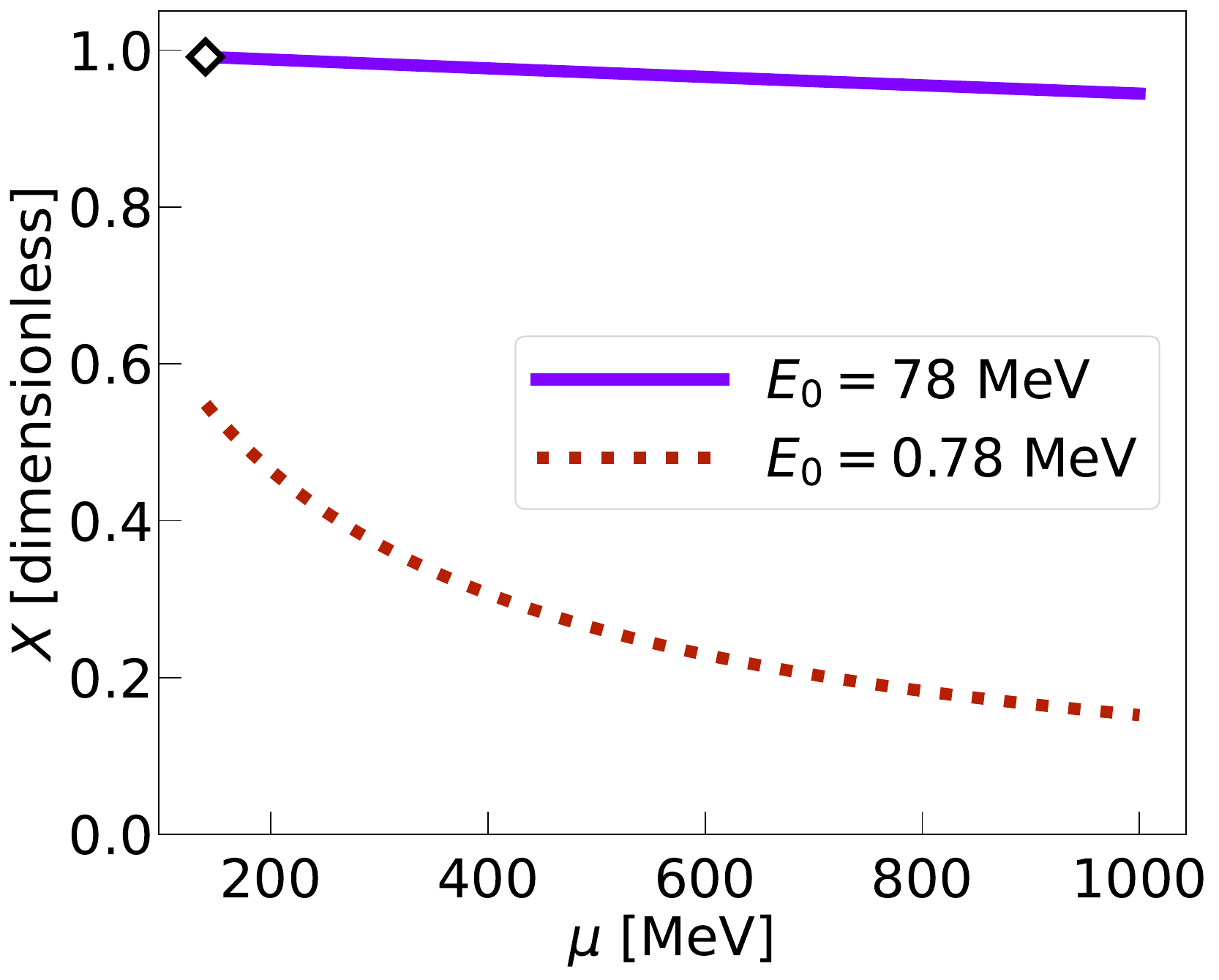}
    \caption{Compositeness $X$ as a function of the cutoff $\mu$. Solid (dotted) line corresponds to $E_{0}=78$ MeV ($E_{0}=0.78$ MeV). Reference value in Table~\ref{tab:param_X3872} is indicated by the diamond. }
    \label{fig:X_mu}
\end{figure}

The dependence of the compositeness $X$ on the cutoff parameter $\mu$ is shown in Fig.~\ref{fig:X_mu}. The cutoff is varied from the reference value $\mu = m_\pi$ given in Table~\ref{tab:param_X3872} up to a typical hadronic energy scale, $\mu = 1000$~MeV. As in the discussion of the binding energy $B$ dependence, the change in $X$ is modest for the reference value of the bare energy, $E_0 = 78$~MeV, where the contribution from the bare state is small. To illustrate the effect of a more significant bare-state component, we also show the result for $E_0 = 0.78$~MeV as the dotted line in the figure. In both cases, the compositeness decreases as the cutoff $\mu$ increases. The decrease is more pronounced for $E_0 = 0.78$~MeV, where the bare-state contribution is enhanced. This behavior is qualitatively different from the results in Refs.~\cite{Song:2022yvz,Kinugawa:2023fbf}, where an increase in compositeness was observed with increasing cutoff when using a sharp cutoff scheme. Thus, we find that the use of a monopole form factor leads to qualitatively different behavior.

\begin{figure}[tbp]
    \centering
    \includegraphics[width=1\linewidth]{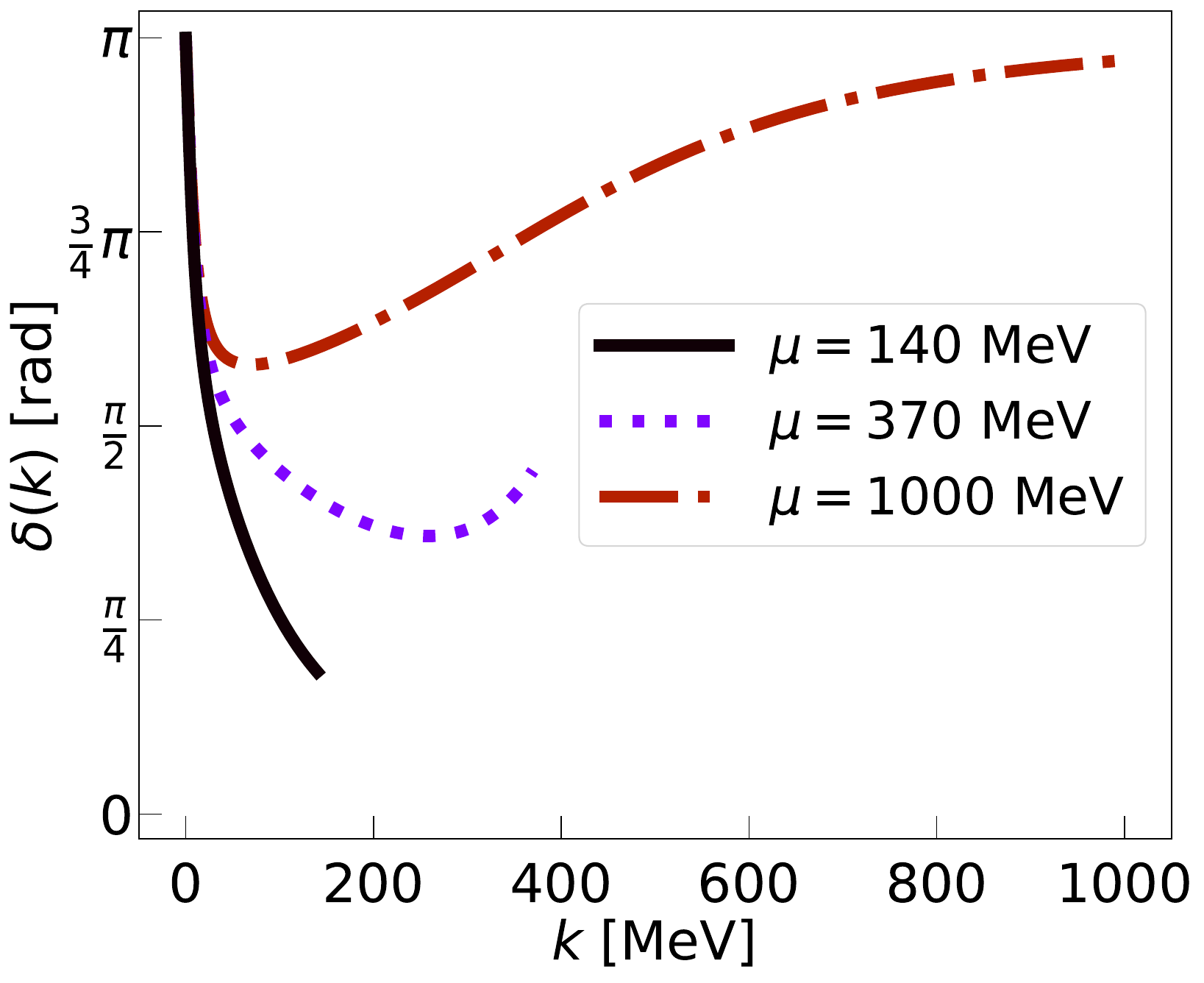}
    \caption{Phase shifts $\delta$ as functions of $k$ for $\mu = 140$~MeV (reference value, solid line), $\mu = 370$~MeV (dashed line), and $\mu = 1000$~MeV (dot-dashed line). The result is plotted up to $k=\mu$ for each $\mu$. The bare energy is set at the  reference value, $E_{0}=78$ MeV. }
    \label{fig:delta_mu}
\end{figure}

Figure~\ref{fig:delta_mu} shows the phase shifts as functions of the momentum $k$ for $\mu = m_\pi = 140$~MeV, $\mu = 370$~MeV, and $\mu = 1000$~MeV. Here we plot $\delta$ as functions of $k$ up to the cutoff scale $\mu$. We take the reference value $E_0 = 78$~MeV for the bare energy, and for each $\mu$, the phase shift is plotted up to $k = \mu$. We observe that the phase shifts near the threshold, where $k$ is small, are relatively insensitive to the variation of $\mu$. However, as the momentum increases, the differences in the phase shifts become more pronounced. This can be understood as follows: since the coupling constant is adjusted to reproduce the same binding energy for each $\mu$, the low-energy behavior of the phase shift near the threshold remains unchanged. At higher momenta, however, the difference in the cutoff scale affects the momentum dependence of the interaction, leading to visible deviations in the phase shift.

\begin{figure}[tbp]
    \centering
    \includegraphics[width=1\linewidth]{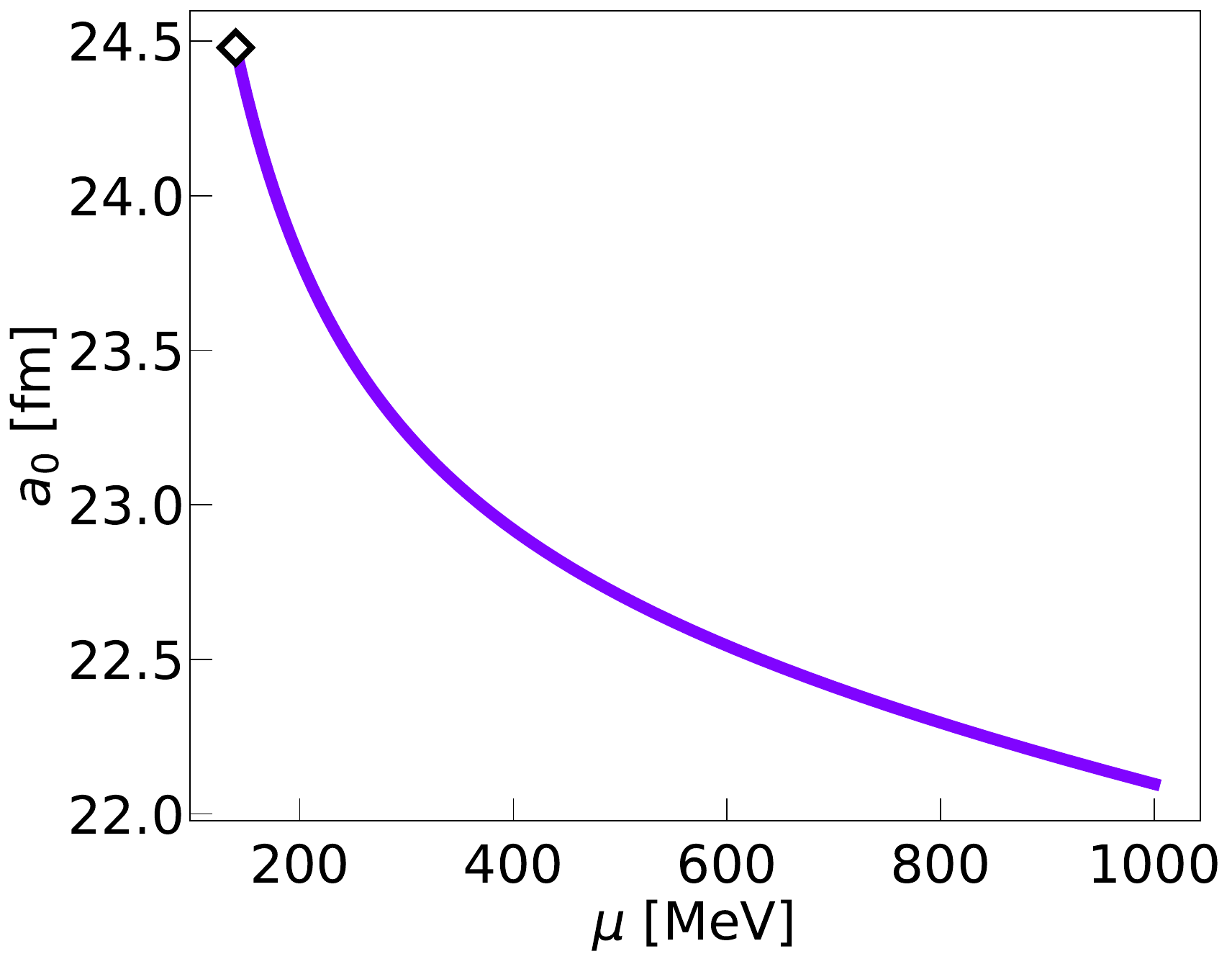}
    \caption{Scattering length $a_0$ as a function of the cutoff $\mu$. Reference value in Table~\ref{tab:param_X3872} is indicated by the diamond.}
    \label{fig:a0_mu}
\end{figure}

\begin{figure}[tbp]
    \centering
    \includegraphics[width=1\linewidth]{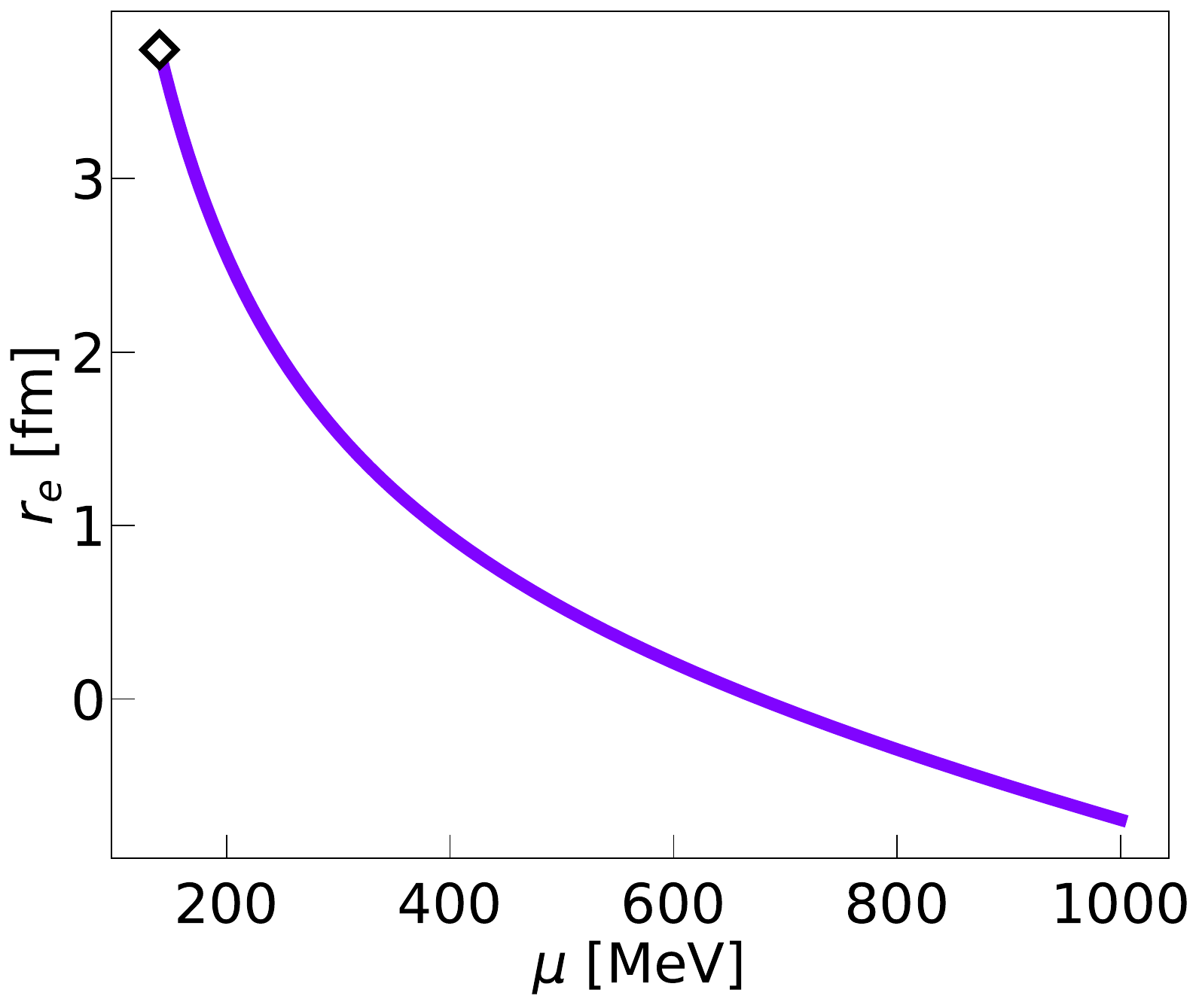}
    \caption{Effective range $r_e$ as a function of the cutoff $\mu$. Reference value in Table~\ref{tab:param_X3872} is indicated by the diamond.}
    \label{fig:re_mu}
\end{figure}

Figures~\ref{fig:a0_mu} and \ref{fig:re_mu} show the scattering length $a_0$ and the effective range $r_e$, respectively, as functions of the cutoff parameter $\mu$ varied within the range $140\leq \mu \leq 1000$~MeV. Again, the bare energy is set as $E_0 = 78$~MeV. Both $a_0$ and $r_e$ decrease with increasing $\mu$, but the changes are limited to a few femtometers in magnitude. This mild variation can be attributed to the facts that the binding energy is held fixed and the associated change in compositeness $X$ remains small in this range. From the weak-binding relations~\eqref{eq:a0WBR} and \eqref{eq:reWBR}, it is evident that $a_0$ and $r_e$ are not significantly modified unless either $X$ or the length scale $R = 1/\sqrt{2mB}$ is largely changed. The insensitivity of the phase shift near the threshold to the cutoff variation, as shown in Fig.~\ref{fig:X_mu}, is thus also consistent with the behavior of $a_0$ and $r_e$ under changes in $\mu$.

\subsection{Variation of direct interaction $\omega^h$}

Finally, we investigate the dependence on the strength of the direct interaction, $\omega^{h}$. As shown in Eq.~\eqref{eq:omega_h_lim}, there exists a lower bound $\omega_{b}^{h}$ for attractive interactions ($\omega^{h} < 0$), below which a bound state is formed by the direct interaction alone. While there is no strict upper bound for repulsive interactions in principle, we take the same absolute value as the lower bound to define the upper limit. Thus, we vary $\omega^{h}$ within the range $\omega_{b}^{h} < \omega^{h} < |\omega_{b}^{h}|$ in this analysis.

\begin{figure}[tbp]
    \centering
    \includegraphics[width=1\linewidth]{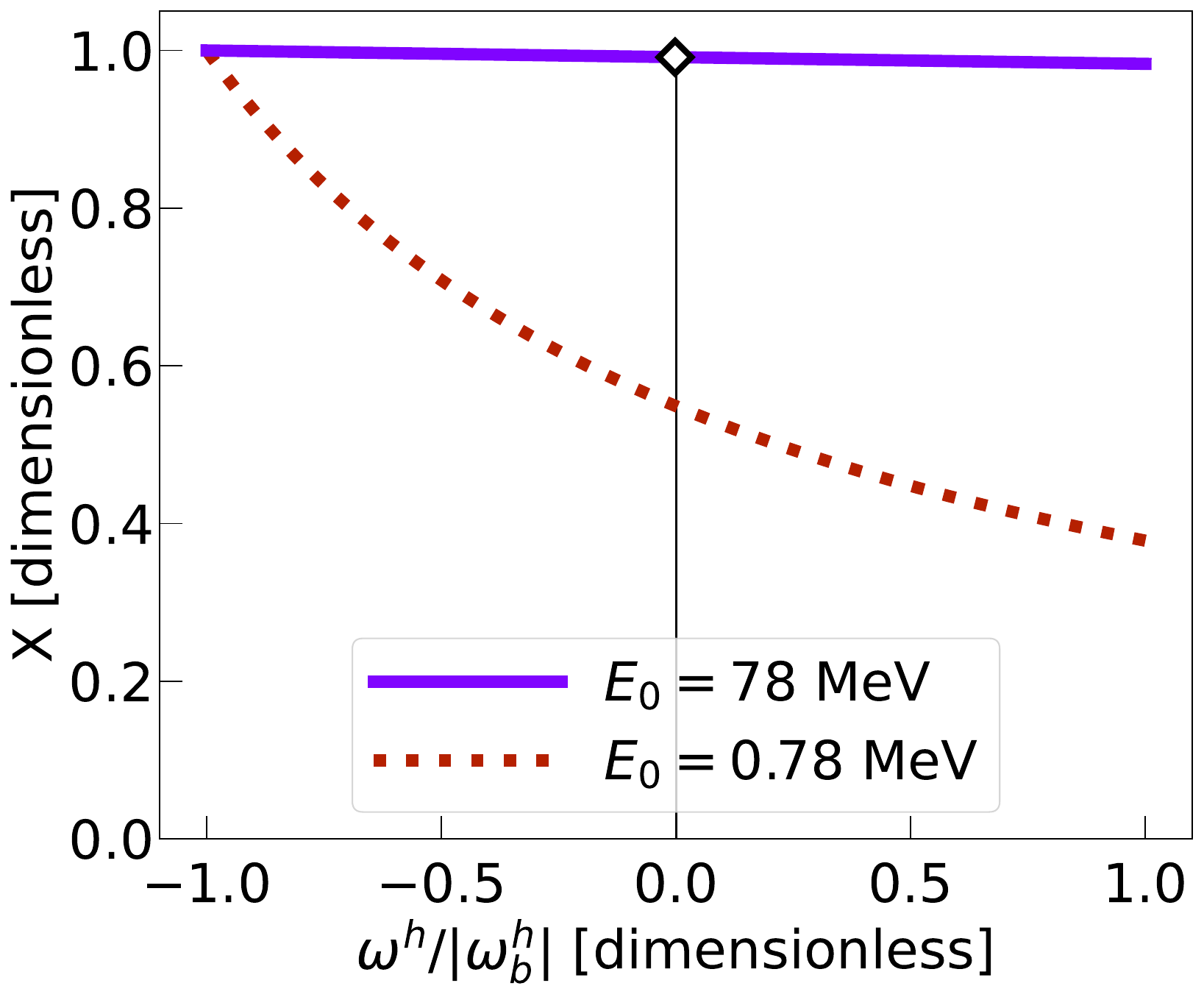}
    \caption{Compositeness $X$ as a function of the normalized direct interaction $\omega^{h}/|\omega^h_b|$. Solid (dotted) line corresponds to $E_{0}=78$ MeV ($E_{0}=0.78$ MeV). Reference value in Table~\ref{tab:param_X3872} is indicated by the diamond. }
    \label{fig:X_omega}
\end{figure}

In Fig.~\ref{fig:X_omega}, the results for the reference value of $E_{0}$ are shown as the solid line, where the compositeness exhibits only a minor variation. As in the case of the cutoff $\mu$, we also show the results for $E_{0} = 0.78$~MeV with a dashed line to enhance the variation in the compositeness. As seen in the figure, increasing $\omega^{h}$, i.e., shifting the direct interaction toward the repulsive side, leads to a decrease in the compositeness. When the binding energy is fixed, the total interaction strength at $E = -B$ is given by $\omega(-B) = \omega^{h} + \omega^{q}(-B)$ and remains constant. Therefore, increasing $\omega^{h}$ implies a decrease in $\omega^{q}(-B)$, which is realized through a change in the coupling strength $g_{0}^{2}$. Since the hadronic component of the interaction is induced by $\omega^{h}$, while the quark-originated elementary component is induced by $\omega^{q}$, the repulsive shift in $\omega^{h}$ enhances the elementary contribution and thus reduces the compositeness. It is also seen that in the limit $\omega^{h} \to \omega_{b}^{h}$ (left edge of the figure), the coupling strength vanishes as $g_{0}^{2} \to 0$, leading to $X = 1$. Note that this also happens for $E_0=0.78$ MeV where $X=0.55$ at $\omega^h=0$. This result confirms the analysis in Ref.~\cite{Kinugawa:2023fbf} by an effective field theory model.

\begin{figure}[tbp]
    \centering
    \includegraphics[width=1\linewidth]{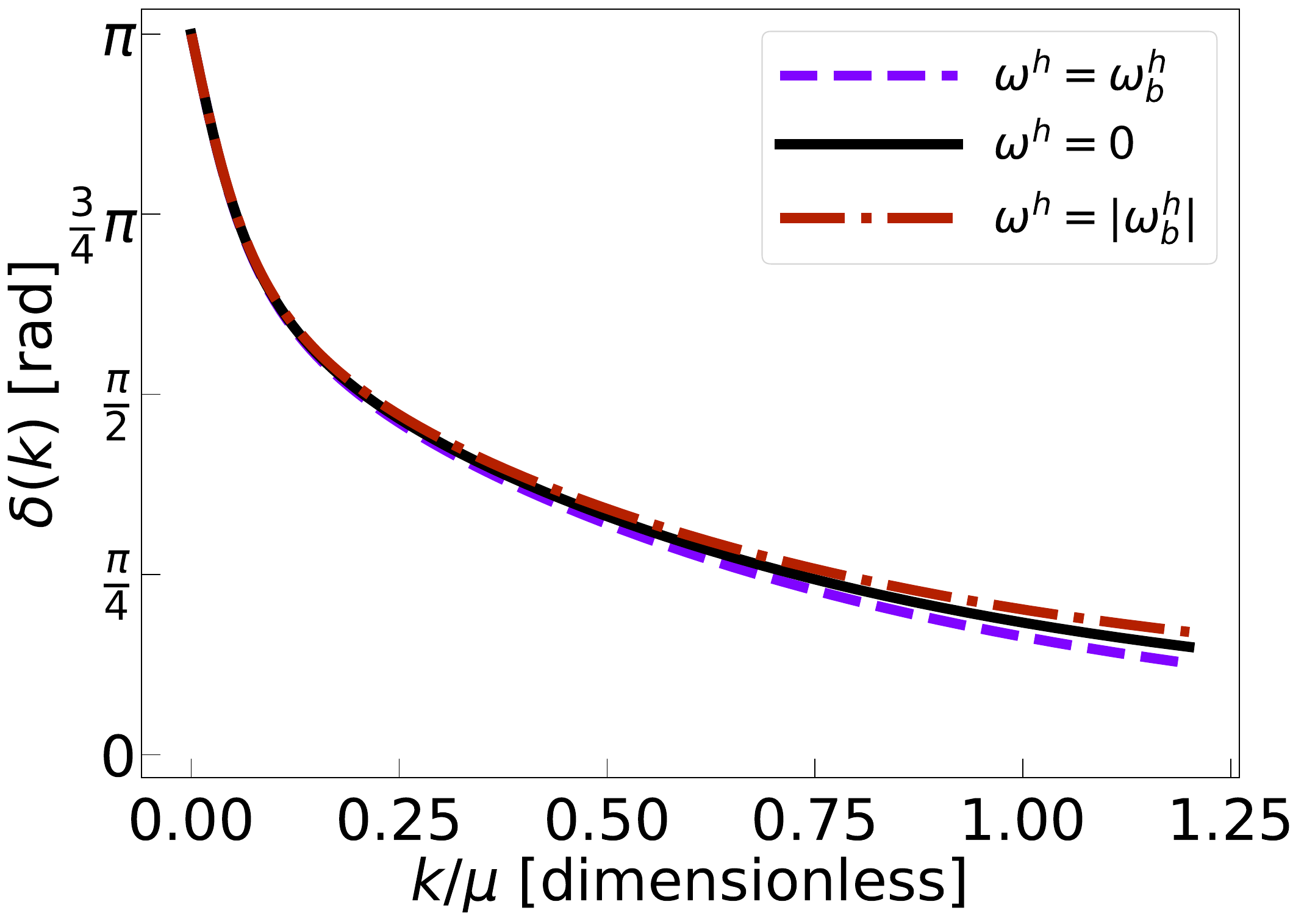}
    \caption{Phase shifts $\delta$ as functions of $k/\mu$ for $\omega^{h} = 0$ (reference value, solid line), $\omega^{h} = \omega^{h}_{b}$ (dashed line), and $\omega^{h} = |\omega^{h}_{b}|$ (dot-dashed line). The bare energy is set at the reference value, $E_{0}=78$ MeV. }
    \label{fig:delta_omega}
\end{figure}

Figure~\ref{fig:delta_omega} shows the phase shifts for three representative values of the direct interaction strength: $\omega^{h} = 0$ (reference value), $\omega^{h} = \omega^{h}_{b}$ (maximal attraction), and $\omega^{h} = |\omega^{h}_{b}|$ (maximal repulsion). Within the plotted range, the phase shifts exhibit negligible variation. This behavior corresponds to the almost flat compositeness $X$ in Fig.~\ref{fig:X_omega}, independently of the direct interaction $\omega^{h}$.

\begin{figure}[tbp]
    \centering
    \includegraphics[width=1\linewidth]{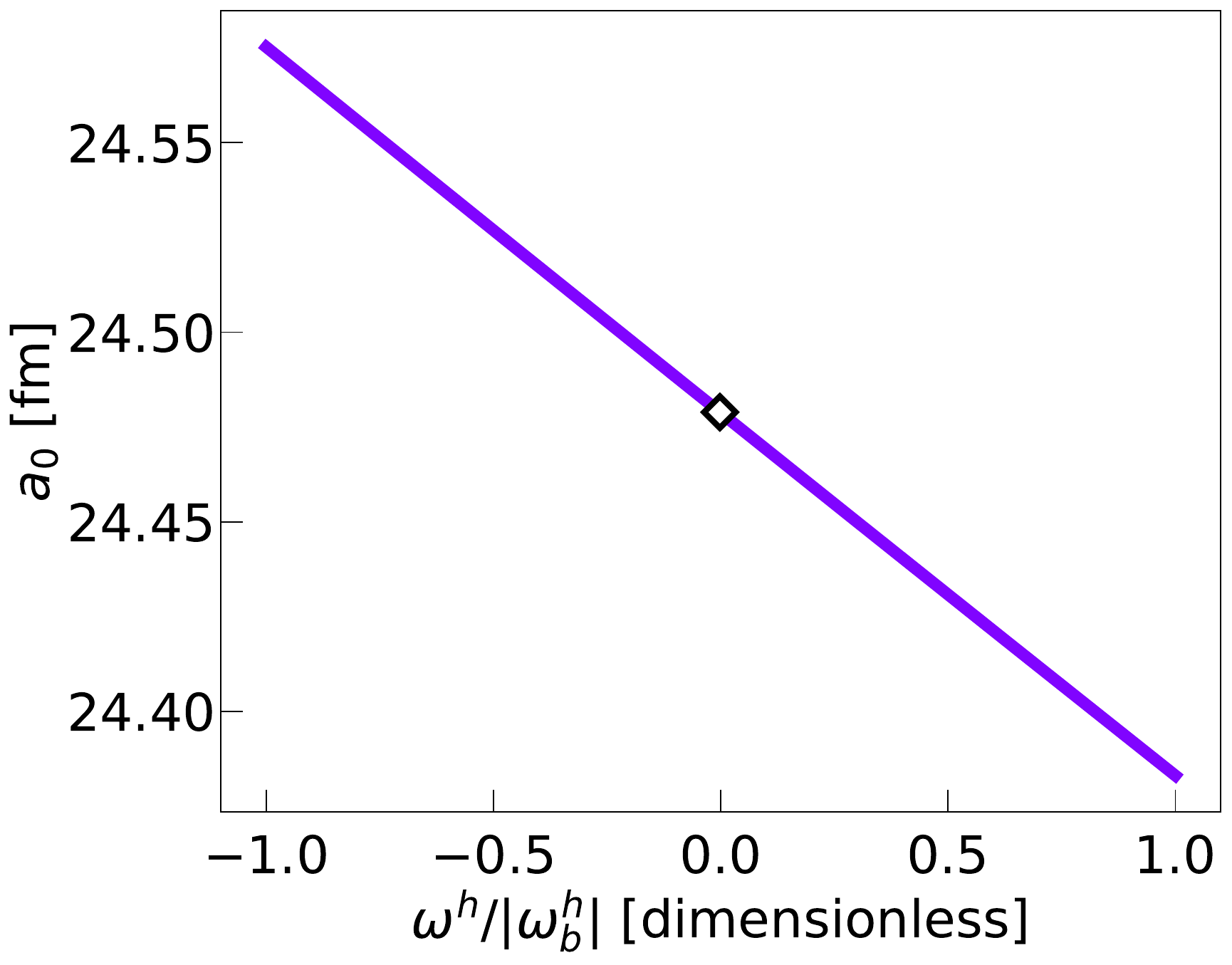}
    \caption{Scattering length $a_0$ as a function of the normalized direct interaction $\omega^{h}/|\omega^h_b|$. Reference value in Table~\ref{tab:param_X3872} is indicated by the diamond.}
    \label{fig:a0_omega}
\end{figure}

\begin{figure}[tbp]
    \centering
    \includegraphics[width=1\linewidth]{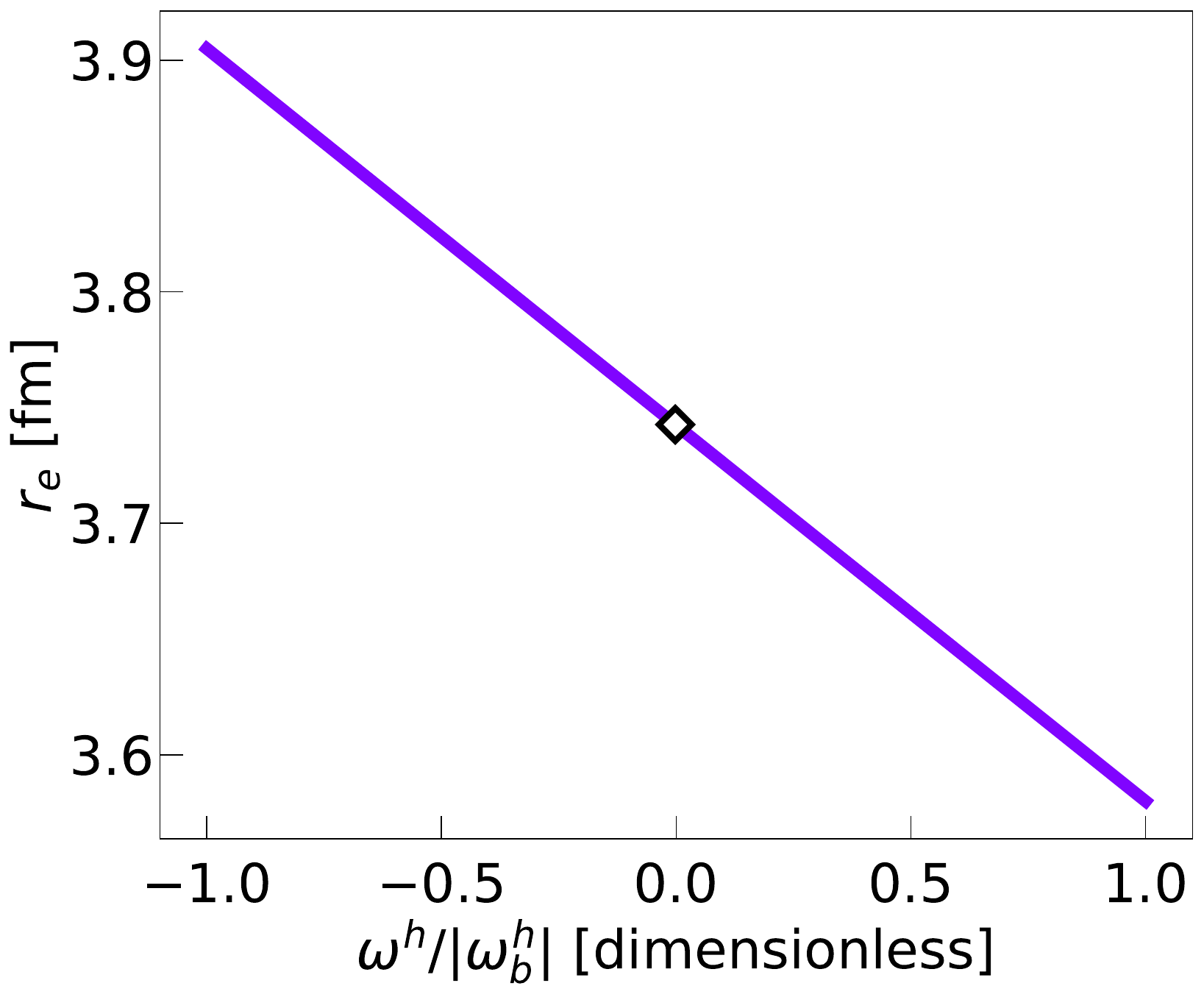}
    \caption{Effective range $r_e$ as a function of the normalized direct interaction $\omega^{h}/|\omega^h_b|$. Reference value in Table~\ref{tab:param_X3872} is indicated by the diamond.}
    \label{fig:re_omega}
\end{figure}

The same results can be observed in the scattering length $a_{0}$ (Fig.~\ref{fig:a0_omega}) and the effective range $r_{e}$ (Fig.~\ref{fig:re_omega}). Both $a_{0}$ and $r_{e}$ decrease as $\omega^{h}$ increases; however, the variations of less than $1$ fm are much smaller compared to the cases with other model parameters. This behavior is consistent with the negligible changes observed in the phase shifts.

\subsection{Effect of local approximation}

In the following, we investigate the effect of the local approximation by computing the phase shifts and wave functions using the local potential obtained via the derivative expansion employed in the HAL QCD method with the choice of $k_0=0$ (see Sec.~\ref{subsec:local}). Specifically, we consider two representative cases: the composite-dominant case with $X = 0.99$, based on the reference parameters listed in Table~\ref{tab:param_X3872}, and the case with a sizable elementary component, $X = 0.55$, realized by choosing $E_{0}=0.78$ MeV. For each case, we analyze the results to assess the validity of the local approximation. As pointed out in Sec.~\ref{subsec:local}, the local potential by the HAL QCD method always gives $X=1$ as shown in Eq.~\eqref{eq:XHAL}. In addition, since the HAL QCD potential is constrained only at $k = 0$, the binding energy of the system can deviate from the exact value under the local approximation.

\subsubsection{Composite-dominant case}

\begin{figure}[tbp]
    \centering
    \includegraphics[width=1\linewidth]{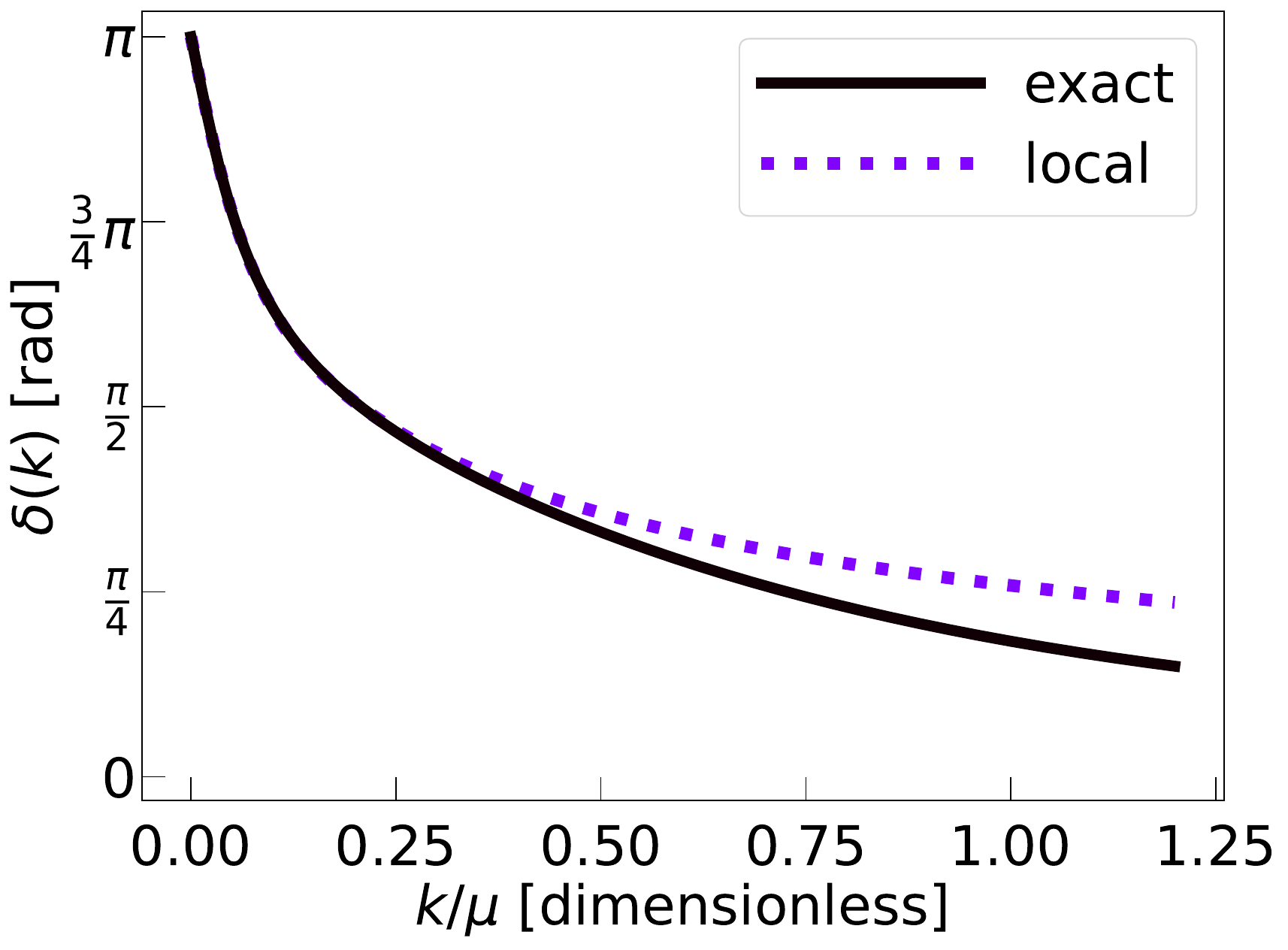}
    \caption{Effect of the local approximation on the phase shift $\delta$ for the composite-dominant case ($X = 0.99$). The solid (dotted) line represents the result of the exact model (local approximation).}
    \label{fig:delta_HAL_1}
\end{figure}

First, we show the results for the composite-dominant case with $X = 0.99$. Figure~\ref{fig:delta_HAL_1} compares the exact phase shift calculated from the nonlocal potential using Eq.~\eqref{eq:kcotdelta} (solid line) with the phase shift obtained numerically from the local potential constructed via the derivative expansion (dashed line). The result shows that the local approximation reproduces the exact phase shift very well, with small deviation in the high-energy region. As the energy increases, small discrepancies gradually appear, as expected from the nature of the local approximation. In the HAL QCD method, the potential is constructed to reproduce the exact phase shift at $k = k_0 = 0$, and thus agreement is guaranteed at low energy. In fact, the scattering length $a_{0}$ is exactly reproduced, and the effective range shows only a small deviation, as summarized in Table~\ref{tab:local}. In this way, the HAL QCD method gives a reasonable local potential which reproduces the exact result, as discussed in Ref.~\cite{Terashima:2023tun}.

\begin{table*}[tbp]
    \centering
    \caption{Effect of the local approximation on the compositeness $X$, binding energy $B$, scattering length $a_0$, and effective range $r_e$. }
\begin{ruledtabular}
    \begin{tabular}{lllllllll}
        $E_0$ (MeV) & $X$    & $X^{\rm local}$ & $B$ (MeV) & $B^{\rm local}$ (MeV) & $a_0$ (fm) & $a_0^{\rm local}$ (fm) & $r_e$ [fm] & $r_e^{\rm local}$ (fm)\\ \hline
        $78$        & $0.99$ & $1$             & 0.040      & 0.040                 & 24.5       & 24.5                   & \phantom{$-1$}3.74       & 3.86 \\
        $0.78$      & $0.55$ & $1$             & 0.040      & 0.083                  & 17.9       & 17.9                   & $-11.6$    & 3.73 
    \end{tabular}
\end{ruledtabular}
    \label{tab:local}
\end{table*}

\begin{figure}[tbp]
    \centering
    \includegraphics[width=1\linewidth]{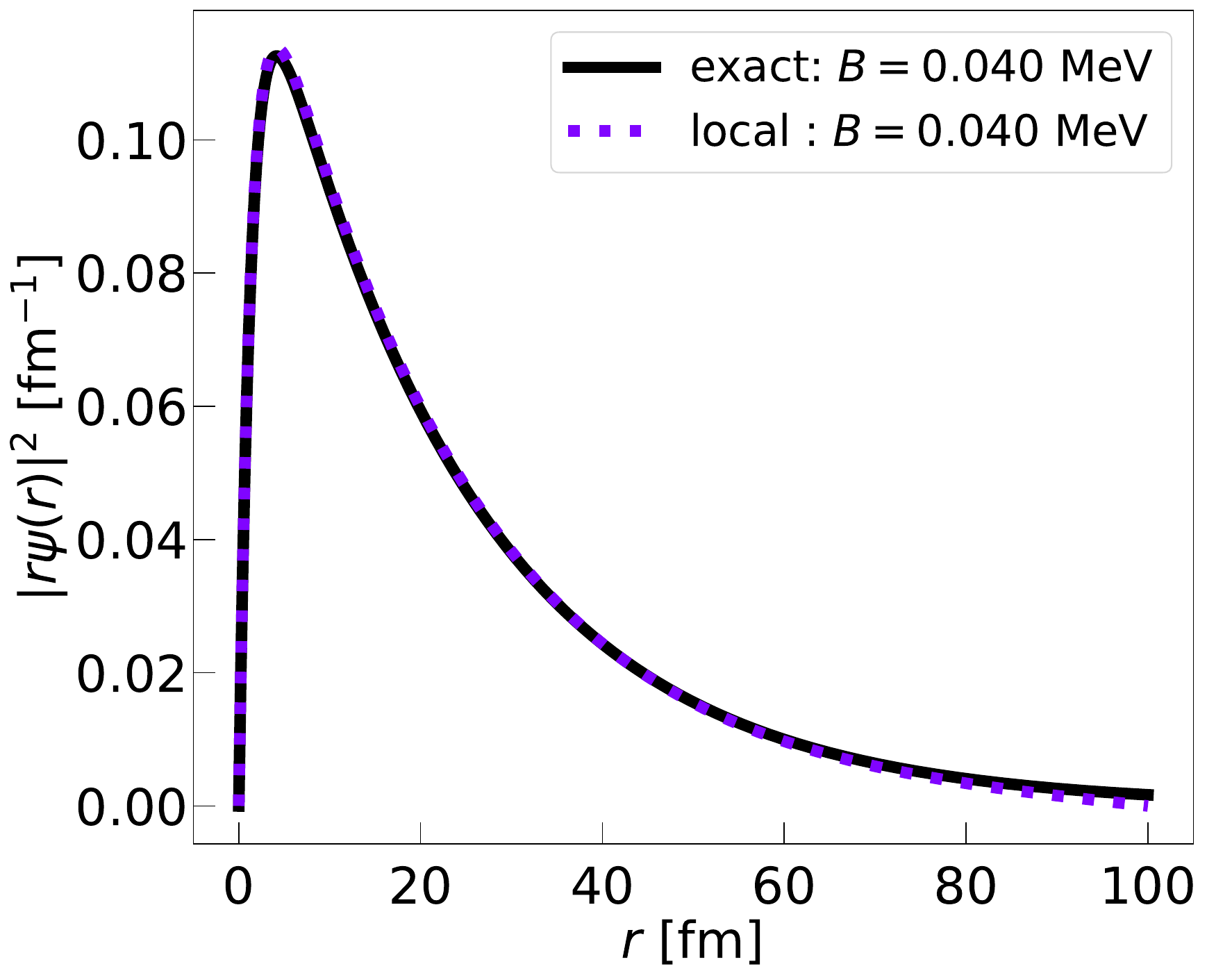}
    \caption{Effect of the local approximation on the wave function $|r\psi(r)|^2$ for the composite-dominant case ($X = 0.99$). The solid (dotted) line represents the result of the exact model (local approximation).}
    \label{fig:w-f_HAL_1}
\end{figure}

Next, we examine the bound-state wave function. As shown in Table~\ref{tab:local}, the binding energy obtained from the local potential derived via the derivative expansion is $B = 0.040$~MeV, which is found to be identical to the exact value at this accuracy. This agreement can be understood by noting that the binding energy lies close to $k = k_0 = 0$, where the local approximation reproduces the interaction accurately. Furthermore, the wave function shown in Fig.~\ref{fig:w-f_HAL_1} is also found to closely match the exact result, indicating that the local approximation captures the spatial structure of the bound state well. These observations suggest that the HAL QCD method not only reproduces scattering observables above the threshold but also provides a good description of the properties of bound states below the threshold.

\subsubsection{Sizable elementary component}

\begin{figure}[tbp]
    \centering
    \includegraphics[width=1\linewidth]{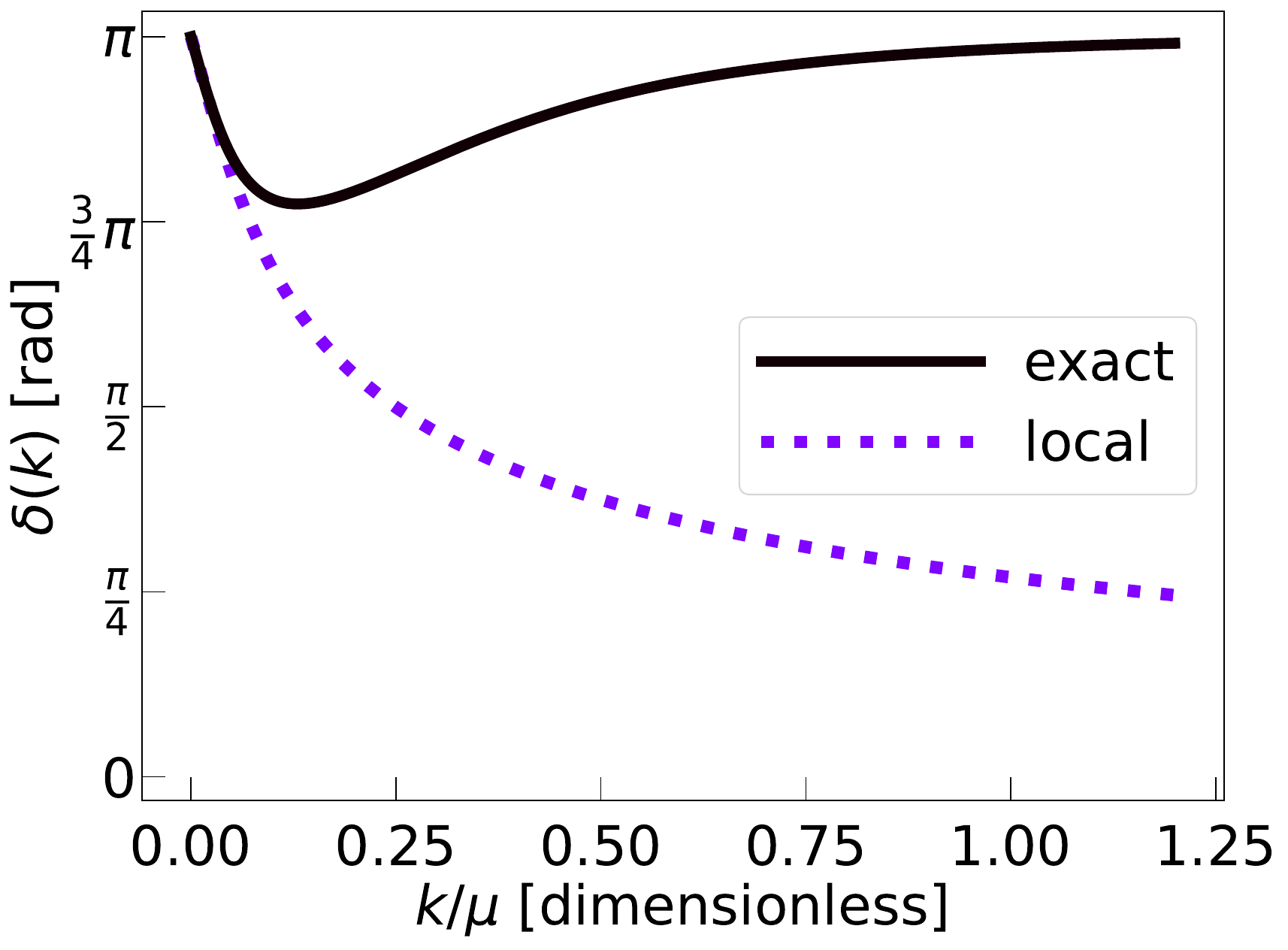}
    \caption{Effect of the local approximation on the phase shift $\delta$ for the case with a sizable elementary component ($X = 0.55$). The solid (dotted) line represents the result of the exact model (local approximation).}
    \label{fig:delta_HAL_0}
\end{figure}

Next, we investigate the local approximation in the case where $E_0 = 0.78$~MeV, corresponding to $X = 0.55$. By construction, the local potential derived from the derivative expansion yields $X = 1$, indicating that the compositeness deviates from the exact value in this case. The comparison of phase shifts is shown in Fig.~\ref{fig:delta_HAL_0}. In contrast to the composite-dominant case, we observe a significant deviation in the phase shifts. This discrepancy can be attributed to the failure of the local approximation to reproduce the correct value of the compositeness. It is also seen in Table~\ref{tab:local} that the effective range deviates notably from the exact result. Owing to the deviation of $X$ from unity, the exact result follows the weak-binding relation [Eq.~\eqref{eq:reWBR}], leading to a large negative value of the effective range. However, the local approximation is constrained as $X = 1$, resulting in an effective-range value nearly unchanged from the composite-dominant case. Consequently, while the slope of the phase shift at the threshold is reproduced due to the agreement in the scattering length, the deviation in the effective range becomes pronounced at higher energies, leading to a mismatch in the phase shifts.

\begin{figure}[tbp]
    \centering
    \includegraphics[width=1\linewidth]{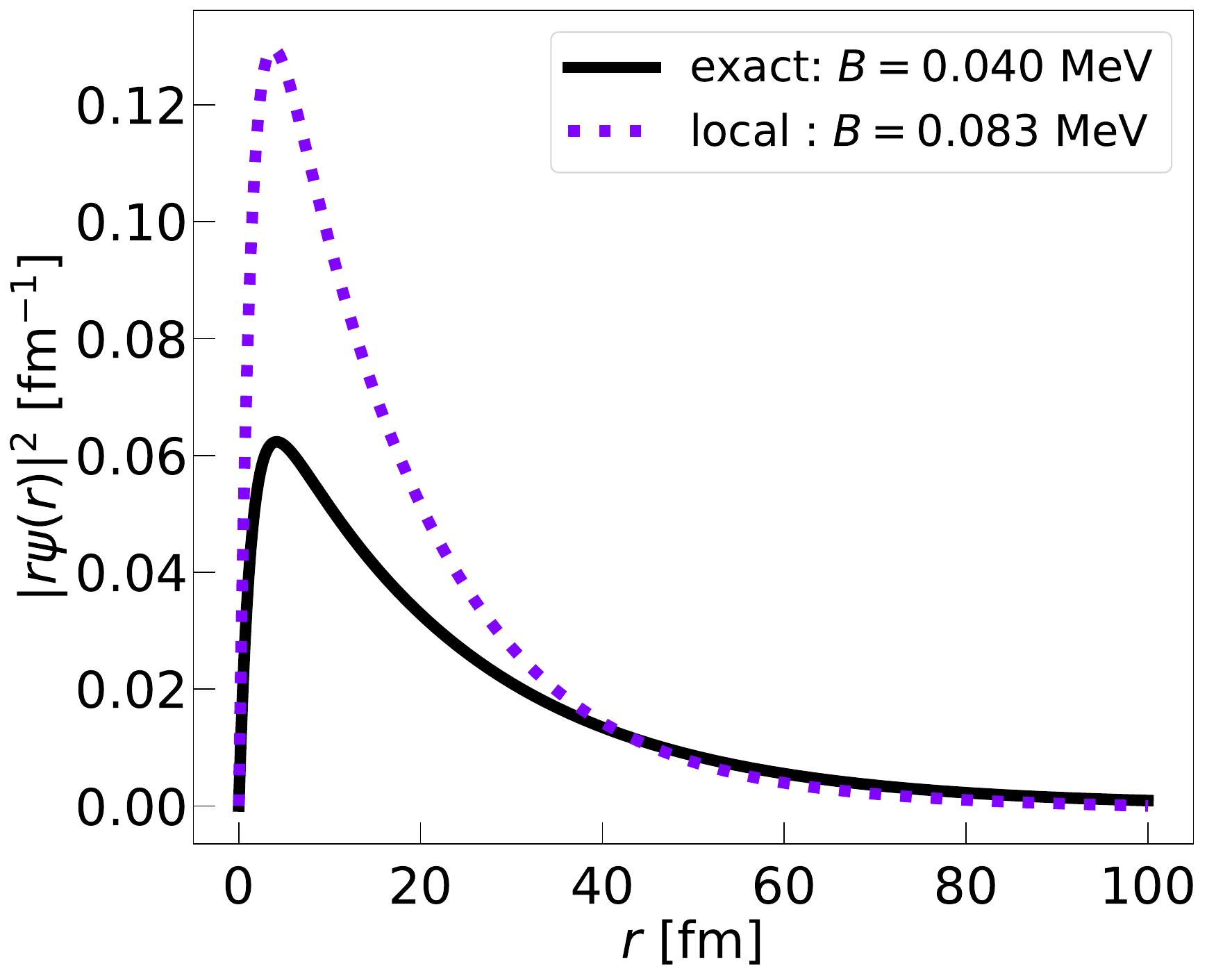}
    \caption{Effect of the local approximation on the wave function $|r\psi(r)|^2$ for the case with a sizable elementary component ($X = 0.55$). The solid (dotted) line represents the result of the exact model (local approximation). }
    \label{fig:w-f_HAL_0}
\end{figure}

Finally, we show the comparison of the wave functions in Fig.~\ref{fig:w-f_HAL_0}. 
For $E_0 = 0.78$~MeV, the binding energy obtained from the local potential is $0.083$~MeV, which is approximately twice the exact value. As a result, the wave function in Fig.~\ref{fig:w-f_HAL_0} becomes more localized near the origin. Furthermore, since the compositeness increases from the exact value $X = 0.55$ to $X = 1$ under the local approximation, the overall magnitude of the wave function also increases from the exact one. 

\section{Application}\label{sec:application}

In this section, we consider the application of the present model to realistic systems. Since our model contains four parameters ($g_{0}$, $E_{0}$, $\mu$, and $\omega^{h}$), the parameter set can be determined by imposing three conditions in addition to the fixed binding energy. Among them, the bare energy $E_{0}$ is estimated based on constituent quark model calculations. The remaining two parameters are fixed by using the scattering length and effective range determined from lattice QCD calculations and experimental analyses. Because the present model does not take into account the decay effects of the obtained bound states, we focus on exotic hadrons with narrow decay widths. In this section, we examine the compositeness of the $X(3872)$, $T_{cc}(3875)^+$, $D_{s0}^*(2317)^{\pm}$, and $D_{s1}(2460)^{\pm}$.

The binding energies $B$ and bare energies $E_{0}$ of the exotic hadrons considered in this study are summarized in Table~\ref{tab:param_exotic}. The ``Channel'' column lists the nearest threshold scattering channels corresponding to each exotic hadron. The binding energy $B$ is evaluated using the central values of the hadron masses reported by the  PDG~\cite{ParticleDataGroup:2024cfk}. For the bare energies $E_{0}$, we adopt the masses of the corresponding $q\bar{q}$ states with the same quantum numbers from the constituent quark model in Ref.~\cite{Godfrey:1985xj} for the $X(3872)$ and $D_{s0}^*(2317)^{\pm}$. For the $D_{s1}(2460)^{\pm}$, we take the lower of the two states predicted in Ref.~\cite{Godfrey:1985xj} at 2530 MeV, which lies closer to the threshold. For the $T_{cc}(3875)^+$, the value of $E_{0}$ is taken from the calculation of the $cc\bar{u}\bar{d}$ state in Ref.~\cite{Karliner:2017qjm}.

\begin{table}[tbp]
    \centering
    \caption{Binding energy $B$ and mass of the quark core state $E_0$ for exotic hadrons, together with the closest hadron-hadron channel which contributes to the molecule component. The binding energy $B$ is determined by the PDG values~\cite{ParticleDataGroup:2024cfk} for the exotic hadron and threshold energy. }
\begin{ruledtabular}
    \begin{tabular}{llcc}
                       & Channel           & $B$ (MeV) & $E_0$ (MeV) \\ \hline
        $X(3872)$      & $D^0\bar{D}^{*0}$ & 0.04 & 78~\cite{Godfrey:1985xj}    \\ 
        $T_{cc}(3875)^+$ & $D^0D^{*+}$       & 0.36 & 6.9~\cite{Godfrey:1985xj}   \\
        $D_{s0}^*(2317)^{\pm}$ & $DK$              & 45   & 117~\cite{Godfrey:1985xj}   \\
        $D_{s1}(2460)^{\pm}$ & $D^*K$            & 45   & 25.8~\cite{Karliner:2017qjm} \\
    \end{tabular}
\end{ruledtabular}
    \label{tab:param_exotic}
\end{table}

The scattering lengths and effective ranges are summarized in Table~\ref{tab:a0re_exotic}. For the $D^{0}\bar{D}^{0}$ channel relevant to the $X(3872)$, we adopt the values compiled in Ref.~\cite{Song:2023pdq}, which are based on the analysis of Ref.~\cite{Baru:2021ldu}. Although Ref.~\cite{Song:2023pdq} does not provide the uncertainty of the scattering length, we assign a conservative theoretical uncertainty of 20\% for the following analysis. For the $D^{0}D^{+}$ channel associated with the $T_{cc}(3875)^+$, we use the results obtained from the experimental analysis reported by the LHCb Collaboration in Ref.~\cite{LHCb:2021auc}. Here, the real part of the complex scattering length extracted from the data is employed. For the $DK$ and $D^{*}K$ channels, corresponding to the $D_{s0}^*(2317)^{\pm}$ and $D_{s1}(2460)^{\pm}$ states, respectively, we take the scattering parameters determined from the lattice QCD analysis in Ref.~\cite{MartinezTorres:2014kpc}.

\begin{table}[tbp]
    \centering
    \caption{Scattering length $a_{0}$ and effective range $r_{e}$ of the scattering channel near the exotic hadrons. }
\begin{ruledtabular}
    \begin{tabular}{lcc}
                       & $a_0$ (fm)         & $r_e$ (fm)  \\ \hline
        $X(3872)$ \cite{Baru:2021ldu,Song:2023pdq} & $28.6\pm 5.7$ & $-2.78< r_e < 1$ \\ 
        $T_{cc}(3875)^+$ \cite{LHCb:2021auc} & $7.16\pm 0.51$  & $-11.9< r_e < 0$  \\
        $D_{s0}^*(2317)^{\pm}$ \cite{MartinezTorres:2014kpc} & $1.3\pm 0.5\pm 0.1$  & $-0.1\pm 0.3\pm 0.1$ \\
        $D_{s1}(2460)^{\pm}$ \cite{MartinezTorres:2014kpc} & $1.1\pm 0.5\pm 0.2$  & $-0.2\pm 0.3\pm 0.1$ \\
    \end{tabular}
\end{ruledtabular}
    \label{tab:a0re_exotic}
\end{table}

\begin{figure}
    \centering
    \includegraphics[width=1\linewidth]{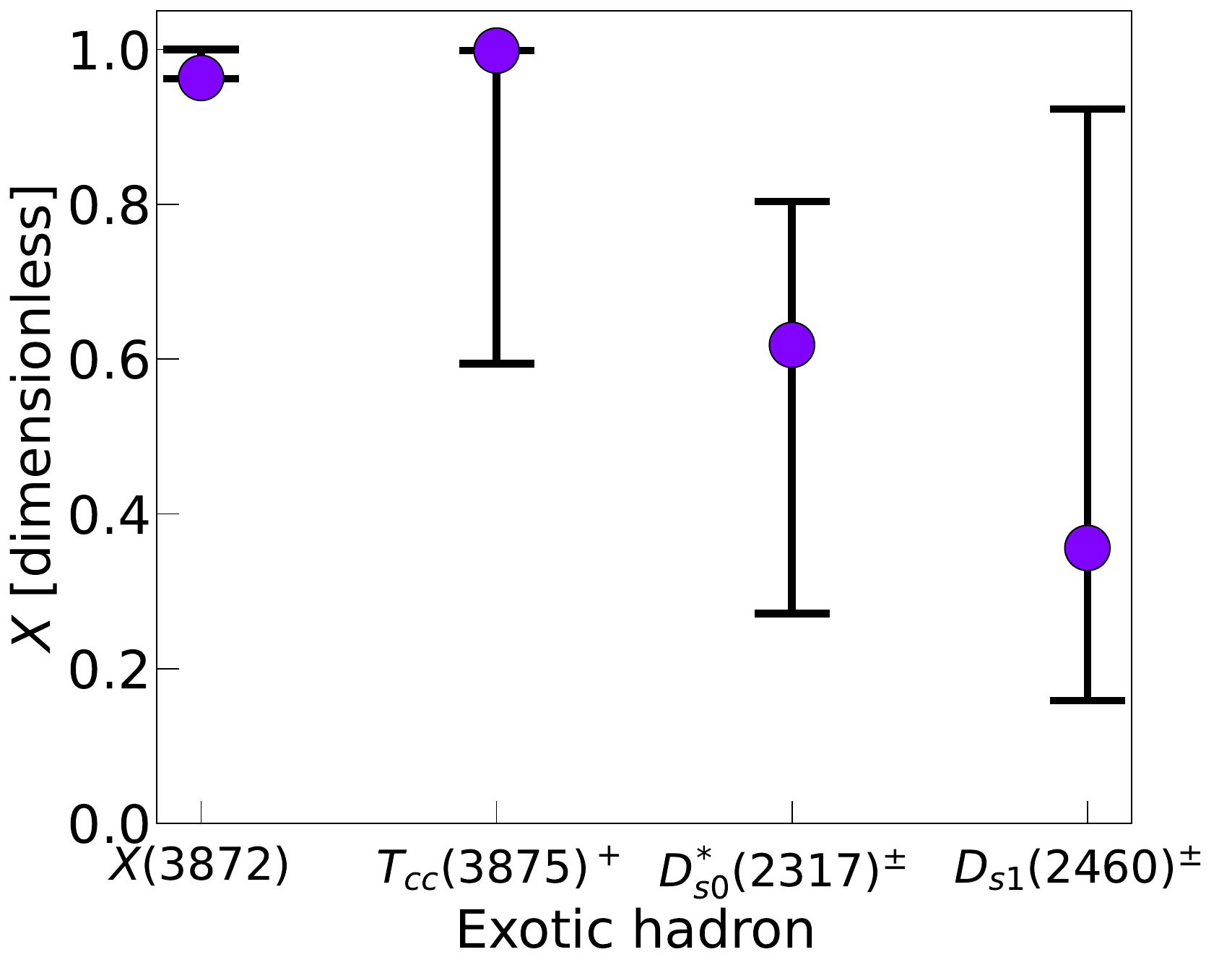}
    \caption{Results of compositeness $X$ of exotic hadrons.}
    \label{fig:result_exotic}
\end{figure}

Now we are in a position to evaluate the compositeness of exotic hadrons. For each exotic hadron, we fix the binding energy $B$ and bare energy $E_{0}$ as in Table~\ref{tab:param_exotic}, vary the cutoff $\mu$ and the strength of the contact interaction $\omega^{h}$ within the reasonable ranges as in Sec.~\ref{sec:results}:
\begin{align}
140  &\leq \mu \leq 1000~{\rm MeV}, \\
-|\omega_b^{h}| &\leq \omega^{h} \leq |\omega_b^{h}|,
\end{align}
and calculate the corresponding scattering length $a_{0}$ and effective range $r_{e}$.
By comparing the obtained values with the empirical ones summarized in Table~\ref{tab:a0re_exotic}, we determine the optimal parameters $\mu$ and $\omega^{h}$ that give the closest results with the central values of $a_{0}$ and $r_{e}$, with which we evaluate the central value of the compositeness $X$. We then identify the range of $\mu$ and $\omega^{h}$ that reproduces $a_{0}$ and $r_{e}$ within their uncertainties, and use the corresponding maximum and minimum values of $X$ to estimate its uncertainty. The resulting compositeness $X$ with errors, together with the corresponding $\mu$ and $\omega^{h}$ values, are summarized in Table~\ref{tab:results_exotic}. The results of the compositeness are also plotted in Fig.~\ref{fig:result_exotic}.

\begin{table}[tbp]
    \centering
\begin{ruledtabular}
    \begin{tabular}{lccc}
                       & $X$ (dimensionless)    & $\mu$ (MeV) & $\omega^h/|\omega^h_b|$ (dimensionless) \\ \hline
        $X(3872)$      & $0.96^{+0.04}_{-0.00}$ & $319$ & $\phantom{-}0.07$ \\ 
        $T_{cc}(3875)^+$ & $1.00^{+0.00}_{-0.40}$ & $726$ & $-0.84$  \\
        $D_{s0}^*(2317)^{\pm}$ & $0.62^{+0.19}_{-0.35}$ & $843$ & $-0.76$ \\
        $D_{s1}(2460)^{\pm}$ & $0.36^{+0.57}_{-0.20}$ & $449$ & $-0.44$ \\
    \end{tabular}
\end{ruledtabular}
    \caption{Results of the compositeness $X$ of exotic hadrons, together with the optimal values of the cutoff $\mu$ and strength of the contact interaction $\omega^{h}$.}
    \label{tab:results_exotic}
\end{table}

From Table~\ref{tab:results_exotic}, the compositeness of the $X(3872)$ is found to be $0.96 \leq  X \leq 1$, indicating that it is almost purely composite dominant. This result is consistent with the analysis of Ref.~\cite{Song:2023pdq}, which shows that the $X(3872)$ is dominantly molecular when the scattering length and effective range are constrained, as well as with other independent evaluations of the compositeness~\cite{Baru:2021ldu,Kinugawa:2022fzn,Wang:2023ovj,Kinugawa:2024crb}. For $T_{cc}(3875)^{+}$, the result also suggests a dominantly composite nature even when uncertainties are taken into account, though the compositeness $X$ exhibits a larger uncertainty than that of the $X(3872)$. Similar tendencies have been reported in other analyses~\cite{Du:2021zzh,Kinugawa:2023fbf,Dai:2023kwv}. Comparing the $X(3872)$ and $T_{cc}(3875)^{+}$, we find that while their threshold and binding energies are almost identical, the bare energy $E_{0}$ differs by an order of magnitude. As discussed in Sec.~\ref{subsec:variationE0}, a smaller $E_{0}$ closer to the threshold generally leads to a larger admixture of the elementary component. Therefore, in the present analysis, the smaller value of $E_{0}$ for $T_{cc}(3875)^{+}$ can be interpreted as the source of the wider uncertainty toward smaller $X$. A similar trend can also be observed when comparing $D_{s0}^*(2317)^{\pm}$ and $D_{s1}(2460)^{\pm}$. As shown in Table~\ref{tab:param_exotic}, their binding energies are of the same order, while the bare energy of $D_{s1}(2460)^{\pm}$ is significantly smaller. Consequently, the compositeness of $D_{s1}(2460)^{\pm}$ is evaluated to be smaller than that of $D_{s0}^*(2317)^{\pm}$, reflecting the stronger admixture of the elementary component.

\section{Summary}\label{sec:summary}

In this study, we have investigated the relationship between the compositeness and physical observables by employing a coupled-channel model that includes both quark and hadronic degrees of freedom. Starting from a coupled-channel Hamiltonian, we naturally describe the bare state appearing in the discussion of compositeness as a bound state in the quark channel. By adopting a Yukawa-type form factor for the transition interaction, we derive analytic expressions for the bound-state condition, compositeness, phase shift, scattering length, and effective range. We also show that the local potential obtained via the derivative expansion in the HAL QCD method always yields a fully composite bound state, since it lacks explicit energy dependence.

In the numerical analysis, we use the model that describes the $X(3872)$ resonance~\cite{Terashima:2023tun} as a reference, and investigate how the compositeness changes under variations of the binding energy and model parameters. Since the reference value of the $X(3872)$ binding energy is very small, representing a weakly bound state, we find that a hadronic molecular state with compositeness close to unity is realized over a wide range of parameters, leading to only minor changes in the corresponding phase shifts. However, by finely tuning the bare energy to be comparable to the binding energy, an elementary-dominant state with compositeness close to zero can be realized. In this case, we have shown that the scattering length becomes small and the effective range becomes large and negative, as expected from the weak-binding relation, resulting in significantly different phase shifts even when the binding energy remains the same.

We show that the local approximation based on the HAL QCD method, applied to the nonlocal potential that includes contributions from quark degrees of freedom, successfully reproduces not only the scattering phase shifts but also the bound-state wave function and compositeness for typical weakly bound states. However, when the contribution from the bare state becomes sizable, we find that the validity of the local approximation is limited to the near-threshold energy region. However, realizing a weakly bound state that is not predominantly composite requires fine tuning of the model parameters, and such a situation is generally considered to be unlikely in nature~\cite{Kinugawa:2023fbf}.

Finally, as an application of the present framework, we evaluate the compositeness of the $X(3872)$, $T_{cc}(3875)^+$, $D_{s0}^*(2317)^{\pm}$, and $D_{s1}(2460)^{\pm}$ by taking into account the scattering length and effective range extracted from experimental analyses and lattice QCD calculations. It is found that the present constraints on these quantities allow us to determine the compositeness with reasonable uncertainties. In particular, the result for the $X(3872)$ indicates that the elementary-dominant solutions, which appear only under fine-tuned parameter conditions, are excluded once the constraints on the scattering length and effective range are imposed. These results demonstrate that the estimation of the bare energy position, together with the constraints on observables such as the scattering length and effective range, plays a crucial role in determining the degree of compositeness.

While this study has clarified the relationship between the compositeness of a near-threshold bound state and the scattering phase shift, in realistic studies of exotic hadrons, direct measurements of two-body scattering phase shifts are rarely available. Instead, experimental information often comes indirectly through observables such as invariant mass distributions in decay processes. Therefore, a natural future direction is to investigate how the compositeness and wave function of bound states affect such observables in high-energy experiments, using the present findings as a theoretical foundation.\\
\vspace{-0.1cm}

\begin{acknowledgments}
    The authors thank Tomona Kinugawa, Feng-Kun Guo, and Eulogio Oset for useful discussions.
    This work has been supported in part by the Grants-in-Aid for Scientific Research from JSPS (Grants
No.~JP23H05439 and 
No.~JP22K03637). 
    This work was supported by JST SPRING, Grant No. JPMJSP2156, by JST, the establishment of university fellowships toward the creation of science technology innovation, Grant No. JPMJFS2139, by the Sasakawa Scientific Research Grant from The Japan Science Society, by Sasakawa Grants for Science Fellows (SGSF), and by MIYAKO-MIRAI Project of Tokyo Metropolitan University.
\end{acknowledgments}

\section*{DATA AVAILABILITY}

The data that support the findings of this article are not
publicly available upon publication because it is not techni-
cally feasible and/or the cost of preparing, depositing, and
hosting the data would be prohibitive within the terms of this
research project. The data are available from the authors upon
reasonable request.

%

\end{document}